\documentclass[twocolumn]{aastex631}

\newcommand{\psr}{PSR J2021+3651}
\newcommand{\pwn}{G75.2+0.1}
\newcommand{\mgro}{MGRO J2019+37}
\newcommand{\fgl}{4FGL J2021.1+3651}
\newcommand{\verone}{VER J2019+368}
\newcommand{\vertwo}{VER J2016+371}
\newcommand{\verlep}{VER J2020+368$^{\ast}$}
\newcommand{\verhad}{VER J2018+367$^{\ast}$}
\newcommand{\ehwc}{eHWC J2019+368}
\newcommand{\lhaaso}{LHAASO J2018+3651}
\newcommand{\chandra}{\textit{Chandra}}
\newcommand{\fermi}{\textit{Fermi}}
\newcommand{\nustar}{\textit{NuSTAR}}

\shorttitle{Hard X-ray observation and multiwavelength study of the Dragonfly nebula}
\shortauthors{Woo et al.}

\graphicspath{{figures/}}
\usepackage{xcolor}
\usepackage{multirow}
\usepackage{tabularx}
\usepackage{soul}

\begin{document}

\title{Hard X-ray observation and multiwavelength study \\ of the PeVatron candidate pulsar wind nebula ``Dragonfly"}

\correspondingauthor{Jooyun Woo}
\email{jw3855@columbia.edu}

\author[0009-0001-6471-1405]{Jooyun Woo}
\affiliation{Columbia Astrophysics Laboratory, 550 West 120th Street, New York, NY 10027, USA}

\author[0000-0002-6389-9012]{Hongjun An}
\affiliation{Department of Astronomy and Space Science, Chungbuk National University, Cheongju, 28644, Republic of Korea}

\author[0000-0003-4679-1058]{Joseph D. Gelfand}
\affiliation{NYU Abu Dhabi, PO Box 129188, Abu Dhabi, United Arab Emirates}

\author[0000-0002-3681-145X]{Charles J. Hailey}
\affiliation{Columbia Astrophysics Laboratory, 550 West 120th Street, New York, NY 10027, USA}

\author[0000-0002-9709-5389]{Kaya Mori}
\affiliation{Columbia Astrophysics Laboratory, 550 West 120th Street, New York, NY 10027, USA}

\author[0000-0002-3223-0754]{Reshmi Mukherjee}
\affiliation{Department of Physics and Astronomy, Barnard College, 3009 Broadway, New York, NY 10027, USA}

\author[0000-0001-6189-7665]{Samar Safi-Harb}
\affiliation{Department of Physics and Astronomy, University of Manitoba, Winnipeg, MB R3T 2N2, Canada}

\author[0000-0001-7380-3144]{Tea Temim}
\affiliation{Princeton University, 4 Ivy Ln, Princeton, NJ 08544, USA}

\begin{abstract}
We studied the PeVatron nature of the pulsar wind nebula G75.2+0.1 (``Dragonfly") as part of our \nustar{} observational campaign of energetic PWNe. The Dragonfly is spatially coincident with \lhaaso{} whose maximum photon energy is 0.27 PeV. We detected a compact (radius $1\arcmin$) inner nebula of the Dragonfly without a spectral break in $3-20$ keV using \nustar{}. A joint analysis of the inner nebula with the archival \chandra{} and XMM-Newton observations yields a power-law spectrum with $\Gamma=1.49\pm0.03$. Synchrotron burnoff is observed from the shrinkage of the NuSTAR nebula at higher energies, from which we infer the magnetic field in the inner nebula of 24 $\mu$G at 3.5 kpc. Our analysis of archival XMM data and 13 years of \fermi{}-LAT data confirms the detection of an extended ($\sim10\arcmin$) outer nebula in $2-6$ keV ($\Gamma=1.82\pm0.03$) and non-detection of a GeV nebula, respectively. Using the VLA, XMM, and HAWC data, we modeled a multi-wavelength spectral energy distribution of the Dragonfly as a leptonic PeVatron. The maximum injected particle energy of 1.4 PeV from our model suggests that the Dragonfly is likely a PeVatron. Our model prediction of the low magnetic field (2.7 $\mu$G) in the outer nebula and recent interaction with the host supernova remnant's reverse shock (4 kyrs ago) align with common features of PeVatron PWNe. The origin of its highly asymmetric morphology, pulsar proper motion, PWN-SNR interaction, and source distance will require further investigations in the future including a multi-wavelength study using radio, X-ray, and gamma-ray observations.
\end{abstract}

\section{Introduction} \label{sec:intro}

Pulsar wind nebulae (PWNe) of energetic (spin-down luminosity $\dot{E}>10^{36}$ erg/s) middle-aged (characteristic age $\tau=10-100$ kyr) pulsars are often associated with very-high-energy (VHE, above 1 TeV) sources (e.g., \citet{HESS18}). Many of them are luminous above a hundred TeV without a hint of a spectral cutoff (e.g., \citet{Abeyesekara20} and \citet{Sudoh21}). Recently, the higher energy regime of their spectra was unveiled by the Large High Altitude Air Shower Observatory (LHAASO), the first gamma-ray observatory sensitive to PeV-energy gamma rays, and their detection of 14 Galactic ultra-high-energy (UHE, above 100 TeV) sources (\citet{Cao21a}, \citet{Cao21b}, and \citet{Cao21c}). The highest photon energies detected from these sources range from several hundred TeV to above 1 PeV: irrefutable evidence of particle acceleration above 1 PeV in both hadronic (neutral pion decay) and leptonic (inverse Compton scattering) cases. Identifying these Galactic ``PeVatrons" is the key to the origin of the highest-energy Galactic cosmic rays observed on the Earth (in hadronic case) as well as a better understanding of the particle acceleration, radiation, and transportation mechanism (in both hadronic and leptonic case).

The majority of the LHAASO sources are spatially coincident with middle-aged energetic PWNe, well-known leptonic particle accelerators. Our \nustar{} observational campaign of energetic PWNe aims to explore the extreme nature of such PWNe (\citet{Mori22}). Broadband hard X-ray observations with \nustar{} provide a unique window to the highest end of their parent particle spectra by resolving their synchrotron radiation without contamination from thermal radiation. Combined with modeling the multi-wavelength (MW) spectral energy distribution (SED) of the PWNe over 20 decades of energy range, it allows deducing the key physical parameters that define the systems, such as the maximum particle energy and magnetic field. Our \nustar{} observation and MW SED modeling have functioned as powerful probes of PWNe as energetic leptonic cosmic ray accelerators in our Galaxy (e.g., \citet{Burgess22}, \citet{Rabbit}, and \citet{K3}).

\pwn{} (``Dragonfly") is one of the eight target PWNe of our \nustar{} observational campaign and is likely associated with \lhaaso{}. The Dragonfly is powered by \psr{} (RA = 20:21:05.40, Dec = +36:51:04.5) first discovered by \citet{Roberts02} as a radio pulsar with a rotation period $P \cong 104$ ms. The radio observation of the pulsar was motivated by the detection of an unidentified X-ray source AX J2021.1+3651 (\citet{Roberts01}), which was a follow-up observation of an unidentified gamma-ray source GeV J2020+3658 (\citet{3EG}). As a middle-aged pulsar whose characteristic age $\tau \equiv P/2\dot{P} \sim 17$ kyr, \psr{} is still energetic, with $\dot{E} \sim 3.4 \times 10^{36} \textrm{ erg s}^{-1}$. \psr{} is detected in X-ray as a soft (mostly) thermal ($kT_{BB} = 0.16\pm0.02$ keV) point source by \chandra{} (\citet{Hessels04} and \citet{Etten08}). The authors of both works reported the detection of X-ray pulsations to be insignificant. The PWN \pwn{} of \psr{} was first observed in X-ray by \citet{Hessels04} and was named the ``Dragonfly" by \citet{Etten08} for its double-torus structure. \fermi{}-LAT (\fermi{}) observation by (\citet{Abdo09}) detected GeV pulsations from \psr{}, yet its spectrum sharply cuts off below 10 GeV with no evidence of higher energy emission from the PWN. 

\psr{} and the Dragonfly are located in the Cygnus region, an active star-forming region. The first TeV gamma-ray source detected in spatial coincidence with \psr{} and the Dragonfly was \mgro{} (\citet{Abdo07}). \mgro{} is the second brightest TeV source in the northern hemisphere after the Crab Nebula and largely extended (circular 2D Gaussian with $\sigma=0.32^{\circ} \pm 0.12^{\circ}$). Since its detection, numerous observations in different wavebands have been carried out as attempts to identify the origin of such high energy emissions. \citet{Roberts08} observed the region with the VLA in radio (20 cm) and the XMM-Newton (XMM) in soft X-rays. Both observations revealed a more comprehensive picture of \pwn{} beyond the substructures seen by \chandra{} $-$ a conical diffuse nebula pivoted at \psr{} that extends out to $\sim 20\arcmin$ (radio) and $\sim 10\arcmin$ (soft X-ray) on the west with decreasing surface brightness. In this work, the entire structure of the PWN is referred to as the Dragonfly.

\citet{Aliu14} used the Very Energetic Radiation Imaging Telescope Array System (VERITAS) to resolve \mgro{} into two separate sources: \verone{} and \vertwo{}. While \vertwo{} is dominated by low-energy (below 1 TeV) emission near a supernova remnant (SNR) CTB 87, \verone{} (RA = 20:19:25, Dec = 36:48:14, elliptical 2D Gaussian with major-axis $\sigma_{maj}=0.34^{\circ} \pm 0.03^{\circ}$ and minor-axis $\sigma_{min}=0.13^{\circ} \pm 0.02^{\circ}$) is bright above 1 TeV. With additional 120 hours of data, \citet{Abeysekara18} reported that \verone{} may be resolved into two source candidates, \verlep{} and \verhad{}. The High-Altitude Water Cherenkov gamma-ray observatory (HAWC) found the high-energy emission from \verone{} to be significant even above 56 TeV and named the source \ehwc{}  (\citet{Abeyesekara20}). Its significant detection above 100 TeV by LHAASO with the maximum photon energy $0.27 \pm 0.02$ PeV (\citet{Cao21a}) confirms that one or more PeVatrons of Galactic origin are present in this region. This extreme Galactic source, namely \lhaaso{}, is spatially coincident with multiple possible cosmic ray accelerators, including a Wolf-Rayet (WR) star WR 141, H II region Sh 2-104, \psr{} and the Dragonfly.

In this work, we aim to evaluate the Dragonfly's potential as a leptonic PeVatron. We report the first hard X-ray observation of the Dragonfly using \nustar{}. We analyze the archival \chandra{} and XMM data and 13 years of \fermi{} data on the Dragonfly. We combine the spectra of the Dragonfly extracted from our analyses with the radio and TeV spectra from the previous works to model the MW SED of the Dragonfly. We discuss the common features of PeVatron PWNe, source distance, and magnetic field.

\section{X-ray data analysis} \label{sec:x-ray}

We analyzed two sets of archival \chandra{} data (Obs ID 8502, 34 ks, 2006 Dec 25, and Obs ID 7603, 60 ks, 2006 Dec 29), one set of archival XMM data (Obs ID 0674050101, 135 ks, 2012 Apr 17), one set of new \nustar{} data (Obs ID 40660004002, 61 ks, 2021 May 19). We processed the \chandra{} data using the \texttt{chandra\_repro} task in \texttt{CIAO 4.13} (\citet{ciao}) and the calibration database \texttt{CALDB 4.9.5}. We processed \nustar{} data using the \texttt{nuproducts} task in \nustar{} Data Analysis Software package (\texttt{NuSTARDAS v2.0.0}) contained within \texttt{HEASOFT 6.28} and the \nustar{} calibration database (\texttt{CALDB version 20210315}). We processed the XMM European Photon Imaging Camera (EPIC) MOS data using the \texttt{emchain} and \texttt{emfilter} tasks in the XMM-Newton Extended Source Analysis Software (\texttt{XMM-ESAS}) package contained within the XMM-Newton Science Analysis System (\texttt{SAS v20.0.0}). The net exposure after removing soft proton (SP) flares is 85 ks. The XMM EPIC pn data was not used since it was obtained in small window mode (one single CCD) and hence is inappropriate for observing a large diffuse nebula that extends over multiple CCDs.

\subsection{Timing analysis} \label{subsec:x-ray_time}

\citet{Hessels04} reported a marginal (3.7$\sigma$) detection of X-ray pulsations in $0.5-3$ keV from \psr{} using a \chandra{} data in continuous-clocking mode and contemporaneous radio ephemeris. The same authors reported significant timing noise and a possibility of large glitches in \psr{}. We attempted to search for hard X-ray pulsations from \psr{} using the \nustar{} data. We applied an astrometric correction on the pulsar position to the cleaned event files using the \chandra{} data analyzed in this work. We applied a barycentric correction to these event files for the corrected pulsar position using the \texttt{barycorr} task in \texttt{NuSTARDAS}. We used \texttt{extractor} to select position- and timing-corrected events within the  $r=30\arcsec$ circular region around \psr{} corresponding to the half-power diameter (HPD) of \nustar{}. We generated binned light curves from the selected events in $3-6$, $6-20$, and $3-20$ keV bands (bin size $=1$ ms) using the timing analysis software \texttt{HENDRICS 7.0} (\citet{Bachetti18}). We used the light curves to create power spectra with the timing analysis software \texttt{Stingray v1.1} (\citet{Huppenkothen19}). No significant frequency features were found. Given the lack of contemporaneous pulsar ephemeris, we performed Z$^2_n$ ($n=2$) searches around the radio pulsar frequency and frequency derivative found by \citet{Roberts02}. This search did not yield a significant detection of pulsations.

\begin{figure}[t!]
\includegraphics[width=0.45\textwidth]{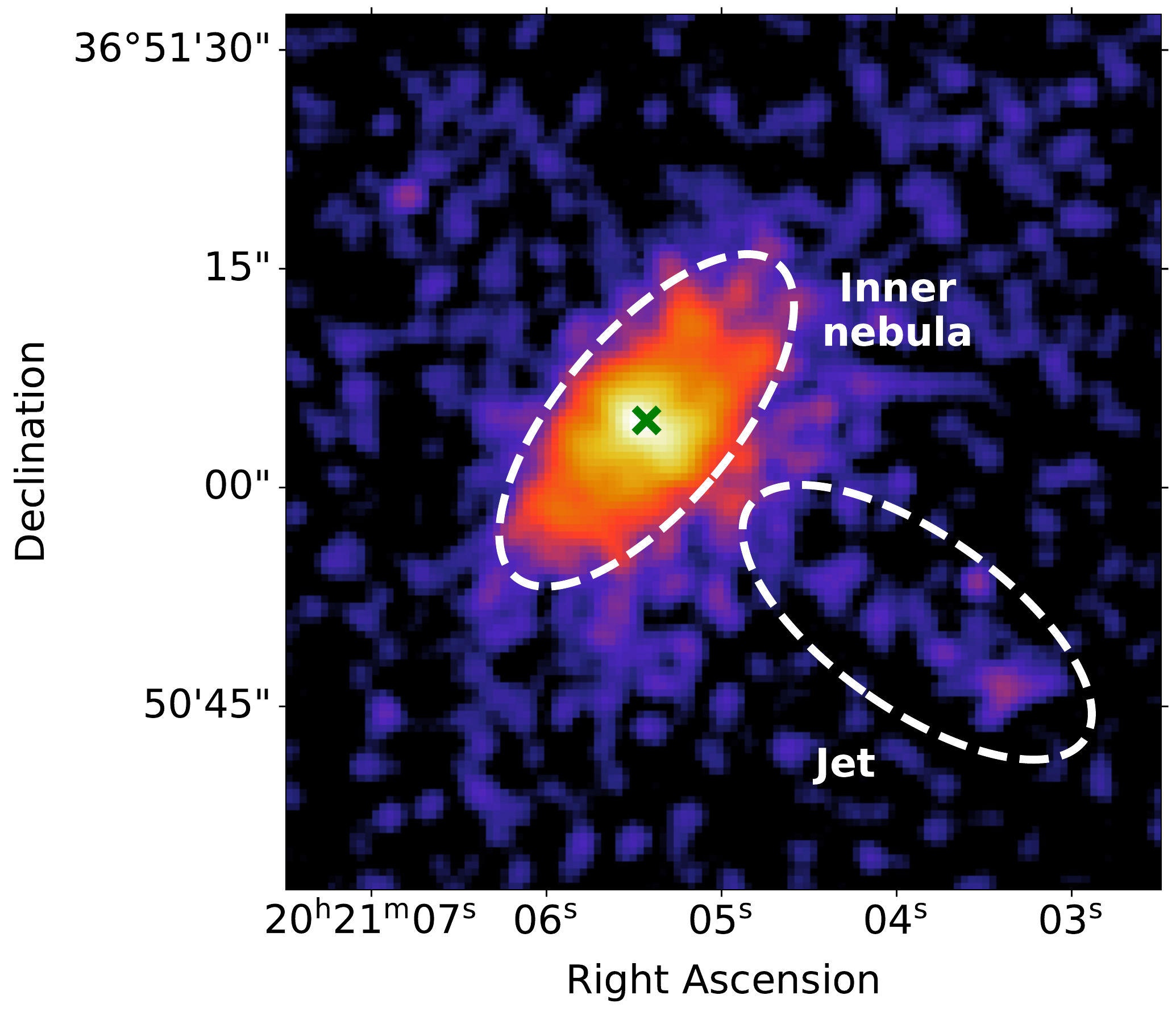}
\includegraphics[width=0.45\textwidth]{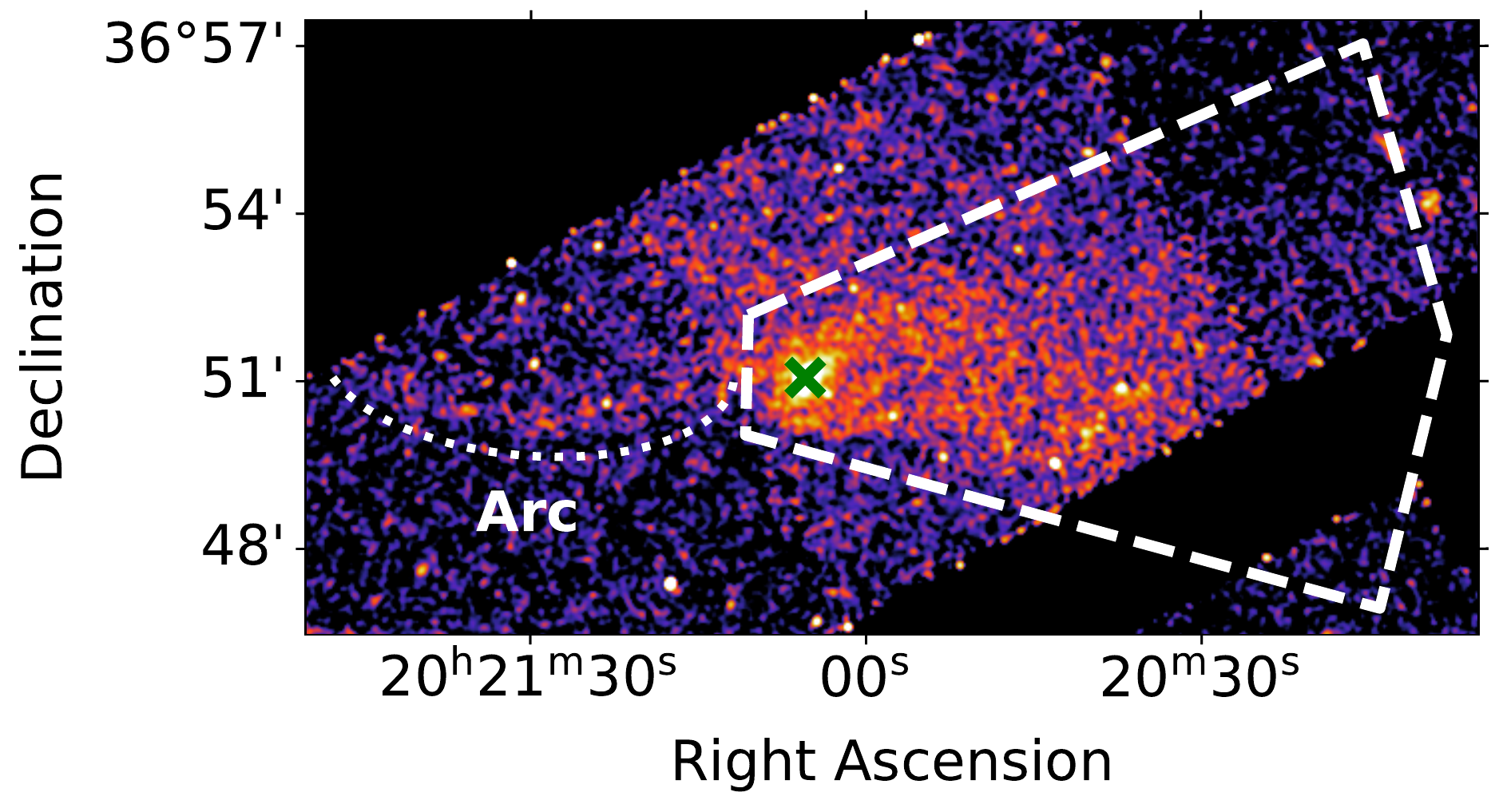}
\caption{Merged (Obs ID 8502 and 7603) and exposure-corrected \chandra{} images of the Dragonfly in $2-6$ keV. The scales were adjusted for better legibility. \psr{} is marked as a cross (X) in both images. \emph{Top}: $1\arcmin\times1\arcmin$ image after Gaussian smoothing with $\sigma=1.5$ pixel $=0.7\arcsec$. The $20\arcsec\times10\arcsec$ inner nebula and the pulsar jet stretching out to $\sim 30\arcsec$ from the pulsar are marked with dashed lines. \emph{Bottom}:  $21\arcmin\times11\arcmin$ image after Gaussian smoothing with $\sigma=3$ pixel $=3.0\arcsec$. The arc in length $\sim7.7\arcmin$ is traced with a dotted line. The extent of the outer nebula seen by XMM is marked as a dashed line.
\label{fig:chandra_image}}
\end{figure}

\subsection{Imaging analysis} \label{subsec:x-ray_image}

\citet{Etten08} resolved the substructures of the \psr{} and the Dragonfly using \chandra{}. Such substructures include pulsar jets, $20\arcsec\times10\arcsec$ double-torus inner nebula, a bow shock standoff, and a peculiar ``arc" stretching toward the east of the pulsar (dotted line in the bottom figure of Figure \ref{fig:chandra_image}). An outer nebula with a size much larger than the inner nebula seen by \chandra{} was discovered by \citet{Roberts08} using XMM. \citet{Zabalza10} used XMM observations covering the region further west to that of \citet{Roberts08} and constrained the size of the outer nebula to $10-15\arcmin$ to the west of \psr{}. \citet{Mizuno17} not only confirmed the western extent of the outer nebula measured by \citet{Zabalza10} using Suzaku but also claimed the emission seen by XMM on the east of \psr{} including the ``arc" is part of the outer nebula.

In this section, we discuss the X-ray morphology of the Dragonfly seen by \chandra{}, XMM, and \nustar{}. We investigate a change in the morphology of the hard X-ray nebula in two different energy ranges: soft band ($3-6$ keV) and hard band ($6-20$ keV). We present the XMM image of the Dragonfly to study the morphology of the outer nebula and briefly discuss the nature of the ``arc." \chandra{} images in $2-6$ keV are compared to the \nustar{} and XMM images in similar energy ranges ($3-6$ keV and $2-6$ keV, respectively). A detailed description of the \chandra{} image can be found in \citet{Etten08}.

\begin{figure}[t!]
\includegraphics[width=0.45\textwidth]{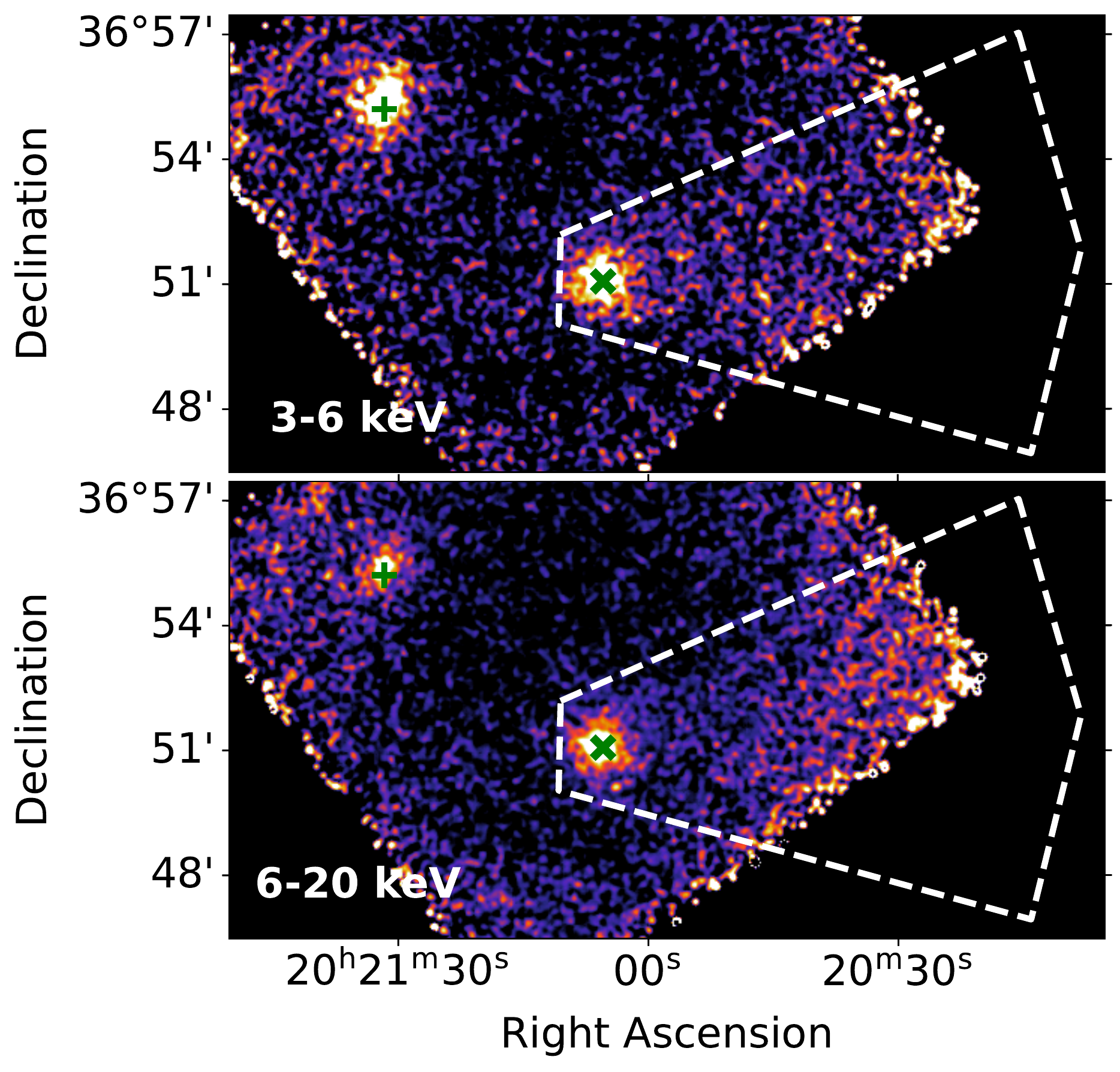}
\caption{Merged (FPMA and FPMB) and exposure- and vignetting-corrected $21\arcmin\times11\arcmin$ \nustar{} images after Gaussian smoothing with $\sigma=1.5$ pixels $=4.7\arcsec$. The scales were adjusted for better legibility. \psr{} and WR 141 are  marked with a cross (X) and a plus (+), respectively. The extent of the outer nebula seen by XMM is marked with a dashed line.
\label{fig:nustar_image1}}
\end{figure}

\subsubsection{Chandra image} \label{subsubsec:chandra_image}

We merged the two \chandra{} observations (Obs ID 8502 and 7603) using the \texttt{merge\_obs} task in \texttt{CIAO} to create an exposure-corrected image in $2-6$ keV. 

Figure \ref{fig:chandra_image} shows smaller (inner nebula and jet) and larger (outer nebula and arc) structures of the Dragonfly. The inner nebula is centered at \psr{} and axis-symmetric along the jet. Its size is $20\arcsec$ along the major axis and $10\arcsec$ along the minor axis in diameter. The jet is measured to extend out to $30\arcsec$ from the pulsar. The observations covered only part of the outer nebula seen by XMM (dashed line in the bottom panel of Figure \ref{fig:chandra_image}), yet it is clearly visible. The arc continues to the edge of the FOV, measuring $7.7\arcmin$ in length.

\begin{figure}[t!]
\includegraphics[width=0.45\textwidth]{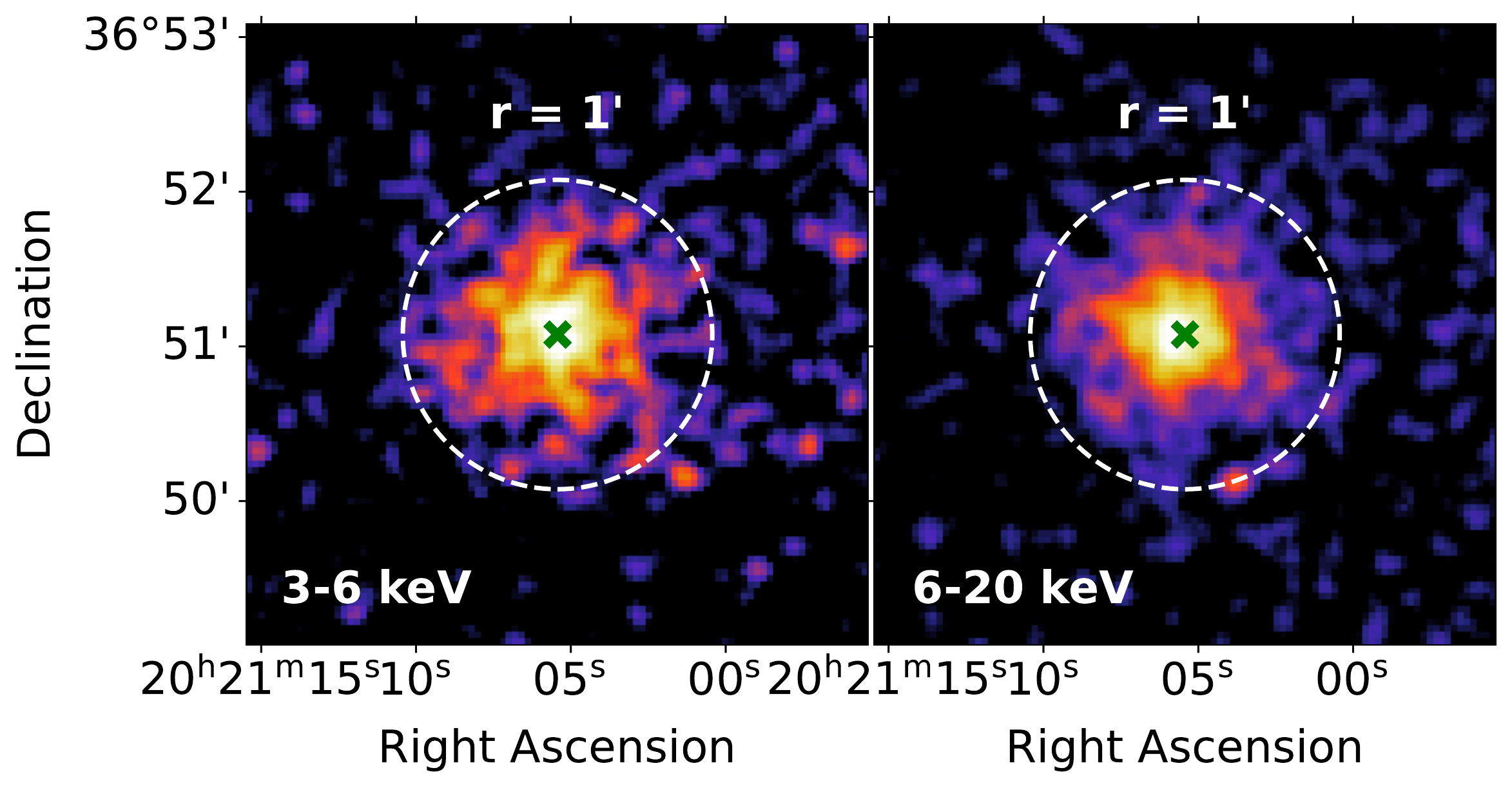}
\caption{$4\arcmin\times4\arcmin$ \nustar{} images of the inner nebula. The images were created following the same procedures described in the caption of Figure \ref{fig:nustar_image1}. \psr{} is marked with a cross (X). A dashed circle of radius $1\arcmin$ is shown as a reference. \label{fig:nustar_image2}}
\end{figure}

\subsubsection{NuSTAR image} \label{subsubsec:nustar_image}

We created images for both focal plane modules (FPMA and FPMB) in the soft band ($3-6$ keV) and the hard band ($6-20$ keV) using \texttt{extractor}. The corresponding exposure maps after vignetting correction were created using \texttt{nuexpomap} task in \texttt{NuSTARDAS}. We combined the FPMA and FPMB images and corrected the exposure using \texttt{XIMAGE} to create Figure \ref{fig:nustar_image1}.

A bright emission is detected in both energy bands at the location of \psr{} (marked with a cross (X) in the figure) and the surrounding region (inner nebula). The west of \psr{} is contaminated by a stray light background, so it is difficult to estimate the emission from the faint outer nebula. WR 141 (marked with a plus sign (+) in the figure) becomes significantly dimmer in the hard band. To examine the detailed morphology of the inner nebula, we created zoomed-in images (Figure \ref{fig:nustar_image2}). The emission is roughly symmetric about the pulsar in both energy bands. The nebula fits well in a radius $1\arcmin$ circle, while it shows an apparent decrease in size in the hard band ($6-20$ keV). We fitted PSF-convolved models to the images using \texttt{Sherpa} (\citet{sherpa}), a fitting and modeling application in \texttt{CIAO}. Both images are fitted with a constant background and a single 2D Gaussian. The FWHM of the Gaussian is $26.5\arcsec \pm 3.2\arcsec$ for the soft band and $15.2\arcsec \pm 2.0\arcsec$ for the hard band.

\begin{figure}[t!]
\includegraphics[width=0.45\textwidth]{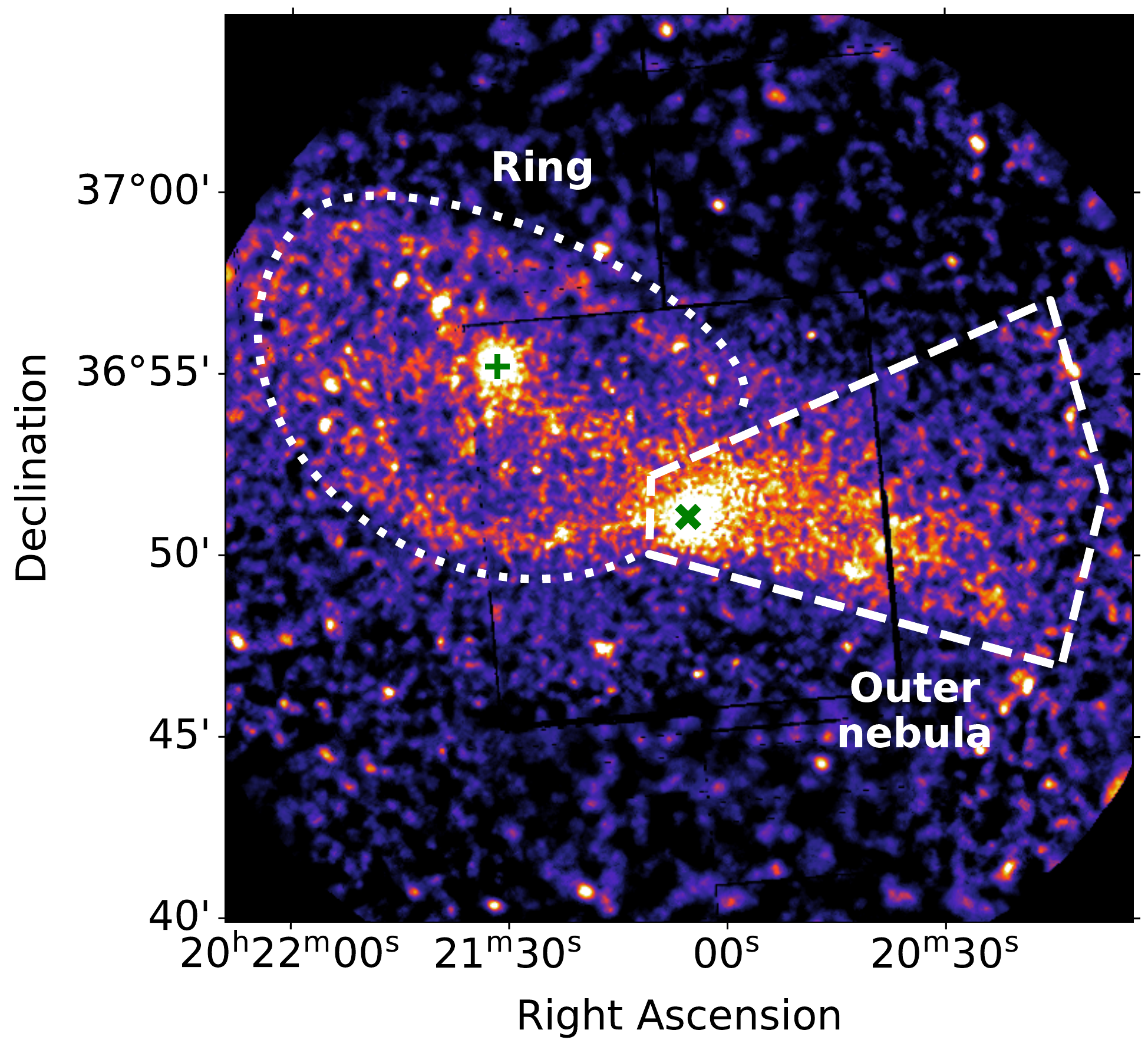}
\caption{Merged (MOS1 and MOS2), QPB-subtracted, exposure-corrected, and smoothed XMM image in $2-6$ keV. \psr{} and WR 141 are marked with a cross (X) and a plus (+) sign, respectively. The extent of the outer nebula $\sim 10\arcmin$ is marked with dashed lines. A ``ring" is marked with a dotted line.
\label{fig:xmm_image}}
\end{figure}

\subsubsection{XMM image} \label{subsubsec:xmm_image}

XMM is the only instrument whose image captures the entirety of the outer nebula in the X-ray band. To study this large diffuse emission, we first removed the contamination of the outer nebula by bright point sources in the FOV, such as \psr{}, WR 141, and a star USNO-B1.0 1268-0448692. We created Swiss cheese masks for MOS1 and MOS2 images that reduce the surface brightness of point sources to 20\% of the surrounding background using the \texttt{cheese} task. These masks were applied to the cleaned event files using the \texttt{mos-spectra} task to create MOS1 and MOS2 images of the entire FOV in $2-6$ keV. The quiescent particle background (QPB) image was generated for the entire FOV in the same energy range using the \texttt{mos\_back} task. Residual SP contamination was found to be negligible (see \S\ref{subsubsec:outer_spec}). No significant instrumental or SWCX background is present in the energy range of our analysis, and no significant stray light background was observed in the image. Therefore, after combining the MOS1 and MOS2 images (\texttt{comb} task), we subtracted only the QPB image, corrected the exposure, and adaptively smoothed it using the \texttt{adapt} task to create Figure \ref{fig:xmm_image}.

Significant emissions are present on the east and west of \psr{} (marked with X). The emission on the east of \psr{} (``ring"-like structure, marked with dotted line) shows no low-energy counterpart in the radio (VLA L band) observation by (\citet{Roberts08}). \citet{Mizuno17} claimed this ring-like structure to be part of the Dragonfly. On the other hand, \citet{Etten08} and \citet{Jin23} detected a bow-shock structure from the inner nebula of the Dragonfly in the X-ray (\chandra{}) and radio (VLA C and L band) observations, respectively. Such detections indicate a supersonic motion of \psr{} toward the east, in which case it is unlikely to expect PWN emission ahead of the bow shock formed by the pulsar. Possible origins of the emission on the west of the pulsar are discussed in the last paragraph of this section.

\begin{figure}[t!]
\includegraphics[width=0.45\textwidth]{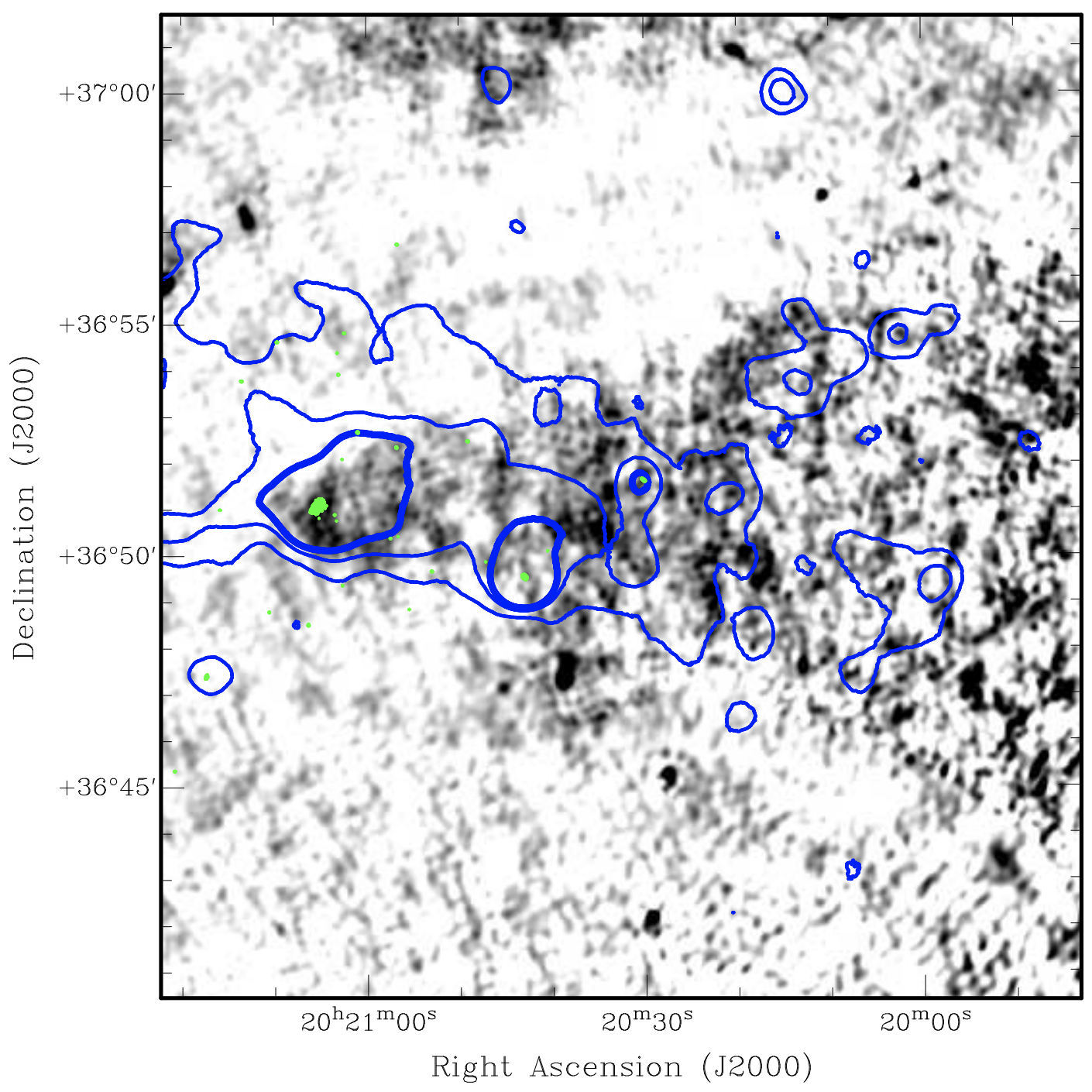}
\caption{VLA 20cm image from \citet{Roberts08}. The radio nebula extends out to $>20\arcmin$ from \psr{}. The XMM contours are overlaid in blue. The permission for the use of the image was acquired from AIP Publishing via RightsLink$^{\textrm{\textregistered}}$.
\label{fig:vla_image}}
\end{figure}

The emission on the west of \psr{} (``outer nebula", marked with dashed line) extends out to $\sim 10\arcmin$ with decreasing surface brightness. This X-ray nebula is spatially coincident with the first half of the radio nebula, as shown in Figure \ref{fig:vla_image}. The radio nebula extends further out to $> 20\arcmin$ (\citet{Roberts08}), whose flux was used for modeling the SED of the Dragonfly (see \S \ref{sec:sed}). For consistency, we analyze the X-ray counterpart of the radio nebula, namely the outer nebula, and use its spectrum for SED modeling.

A ring-like structure is centered at WR 141 and has radius $\sim 5\arcmin$. The ``arc" seen by \chandra{} comprises the lower part of this ring. \citet{Mizuno17} claimed that the ``ring'' is part of the PWN based on the similar spectral index between the ``ring" and the ``outer nebula." \citet{Barkov19} explained the ``arc" as a ``kinetic jet": pulsar wind particles that escaped into the ISM due to magnetic reconnection between the PWN and the ISM and became visible in a high $>$ 10 $\mu$G ISM magnetic field. A similar filamentary emission ahead of the main body of the PWN (``outer nebula") was observed in the ``Snail" PWN, whose ``prongs" may be the result of the interaction between the PWN and the reverse shock of its host SNR (\citet{Temim15}). We propose that the ``ring" in our XMM image is possibly associated with WR 141. WR stars are known to have strong stellar winds that can create a bubble of several parsecs in radius (\citet{Weaver77}). This bubble is often observed as a ring-shaped nebula and can be visible in X-ray (e.g., \citet{Toala12}). The parallax of 0.5024 mas in Gaia DR3 (\citet{GaiaDR3}) implies a 2.0 kpc distance to WR 141. This yields the radius of the bubble = 2.9 pc. Part of the ``ring" was also seen in H$\alpha$ photometry by \citet{Law02}, which the authors postulated to be part of the ring nebula photoionized by WR 141.

\subsection{Spectral analysis} \label{subsec:x-ray_spec}

We present a spectral analysis of \psr{} and its PWN using \chandra{}, XMM, and \nustar{} data. We first characterize the pulsar spectrum with \chandra{} taking advantage of its fine angular resolution (HPD $<0.5\arcsec$). We analyze the spectrum of the inner nebula taking into account the contribution of the pulsar by individually and jointly fitting the \chandra{}, XMM, and \nustar{} spectra. We use the XMM data to study the spectrum of the outer nebula. All the spectral models for X-ray analysis presented in this work were multiplied by a cross-normalization factor (\texttt{const}) to adjust relative normalization between different detectors and instruments.

\subsubsection{Pulsar spectrum} \label{subsubsec:pulsar_spec}

We used \chandra{} data to analyze the spectrum of \psr{}. We extracted the source spectra from a circular region with radius $2\arcsec$ centered at \psr{}, and the background spectra from an annulus around \psr{} with radii $2-4\arcsec$ using the \texttt{specextract} command in \texttt{CIAO}. The source spectra were binned to have at least 3$\sigma$ significance over the background in each bin. We began by fitting an absorbed power law (\texttt{const*tbabs*pow}) to the spectra in $0.5-7$ keV where the source emission dominates over the background. The abundance table was set to \texttt{wilm} (\citet{wilm}) for all the X-ray spectral analyses presented in this work. This model gives a reasonable fit ($\chi^2/d.o.f=130/152$) with the best-fit $\Gamma=1.73^{+0.13}_{-0.12}$ and $N_H=(0.26\pm0.05)\times10^{22}$ cm$^{-2}$. However, this best-fit $N_H$ is $3$ times smaller than the $N_H$ found from the \chandra{} spectra of the inner nebula. When the $N_H$ was fixed to the best-fit value found from the \chandra{} inner nebula spectra ($0.76\times10^{22}$ cm$^{-2}$, see \S \ref{subsubsec:inner_spec}), the fit quality became worse ($\chi^2/d.o.f=190/152$) with much softer $\Gamma=2.87\pm0.15$. This is because the pulsar has significant emission in both below and above 3 keV. Initially, a small $N_H$ was favored to explain the emission below 3 keV. Later, the $\Gamma$ was significantly softened to compensate for the larger $N_H$, leaving the emission above 3 keV poorly fitted. We added a black body component to fit the emission below 3 keV while the power law component explains the emission above 3 keV (\texttt{const*tbabs*(bbod+pow)}). $N_H$ is highly degenerate with the black body temperature and the power law index, so we fixed $N_H$ to the \chandra{} value of the inner nebula. This gives the best-fit $kT=0.13\pm0.01$ keV and $\Gamma=1.63\pm0.17$ with $\chi^2/d.o.f=123/151$. We used this model as the initial pulsar component when jointly fitting the \chandra{} spectra of the inner nebula and the XMM and \nustar{} spectra of the pulsar and the inner nebula (see \S \ref{subsubsec:inner_spec}). We iteratively fit the \chandra{} pulsar spectra by changing the $N_H$ to the best-fit value found from the joint fit of the \chandra{}, XMM, and \nustar{} spectra. The pulsar model converged to $kT=0.11\pm0.01$ keV and $\Gamma=1.77\pm0.17$ with $\chi^2/d.o.f=123/151$. The best-fit parameters are comparable to \citet{Etten08} ($\Gamma=1.73^{+1.15}_{-1.02}$), $kT=0.16\pm0.02$ keV, and $N_H=0.67\times10^{22}$ cm$^{-2}$) considering the degeneracy between $kT$ and $N_H$. The unabsorbed flux of \psr{} in $3-10$ keV is $F_{3-10}=(1.20_{-0.17}^{+0.18})\times10^{-13}$ erg/s/cm$^2$.

\begin{figure}[t!]
\centering
\includegraphics[width=0.45\textwidth]{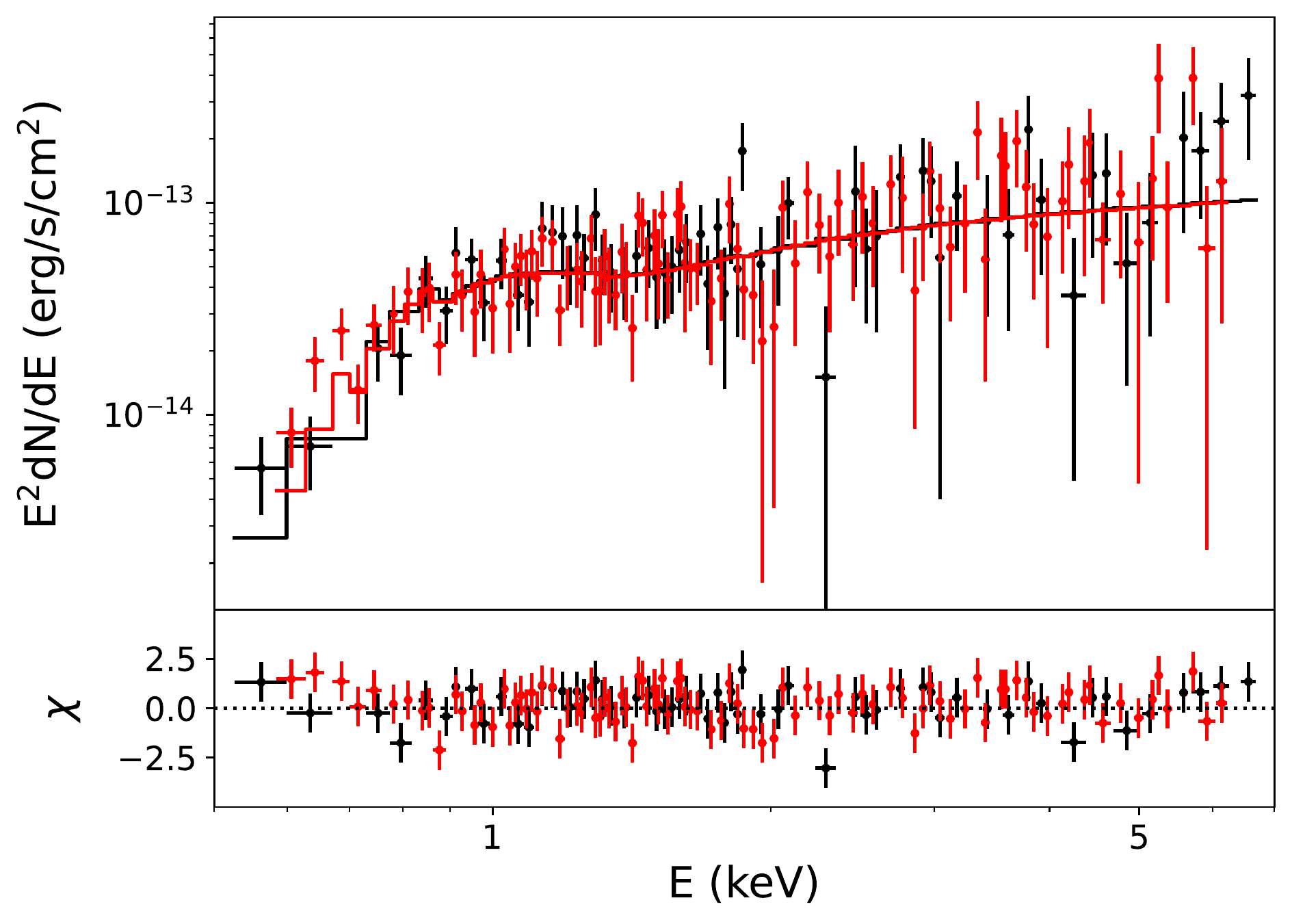}
\caption{\chandra{} ObsID 8502 and 7603 (black and red crosses, respectively) spectra of \psr{}. The background is dominant outside of $0.5-7$ keV. The source spectra were extracted from a circular region with radius $2\arcsec$ centered at the pulsar. The background spectra were extracted from an annulus region with radii $2-4\arcsec$ centered at the pulsar.
\label{fig:x-ray_pulsar_spec}}
\end{figure}

\subsubsection{Inner nebula spectrum} \label{subsubsec:inner_spec}

We individually and jointly fitted the \chandra{}, XMM, and \nustar{} spectra of the inner nebula. The \chandra{} spectra were extracted from an annulus region with radii $2-20\arcsec$ centered at \psr{}. The XMM spectra were  extracted from circular regions with radius $40\arcsec$ using the \texttt{xmmselect} task in \texttt{SAS}. Response files were generated using \texttt{rmfgen} and \texttt{arfgen} tasks. \nustar{} spectra were extracted from circular regions with radius $1\arcmin$ using \texttt{nuproducts} task in \texttt{NuSTARDAS}. The sizes of the source region for different instruments were determined considering the PSF sizes of the instruments ((HPD $1\arcsec$ for \chandra{}, $34\arcsec$ for XMM, and $58\arcsec$ for \nustar{}) and the cross-normalization term between the source spectra. The background spectra for all three instruments were taken from a $2\arcmin\times2\arcmin$ box in a nearby source-free region. Fitting was performed in the energy range where the source emission dominates over the background ($0.5-7$ keV for \chandra{}, $0.5-8$ keV for XMM, $3-20$ keV for \nustar{}). All spectra were binned such that the source counts have at least 3$\sigma$ significance above the background counts in each bin. 

\begin{figure}[t!]
\centering
\includegraphics[width=0.45\textwidth]{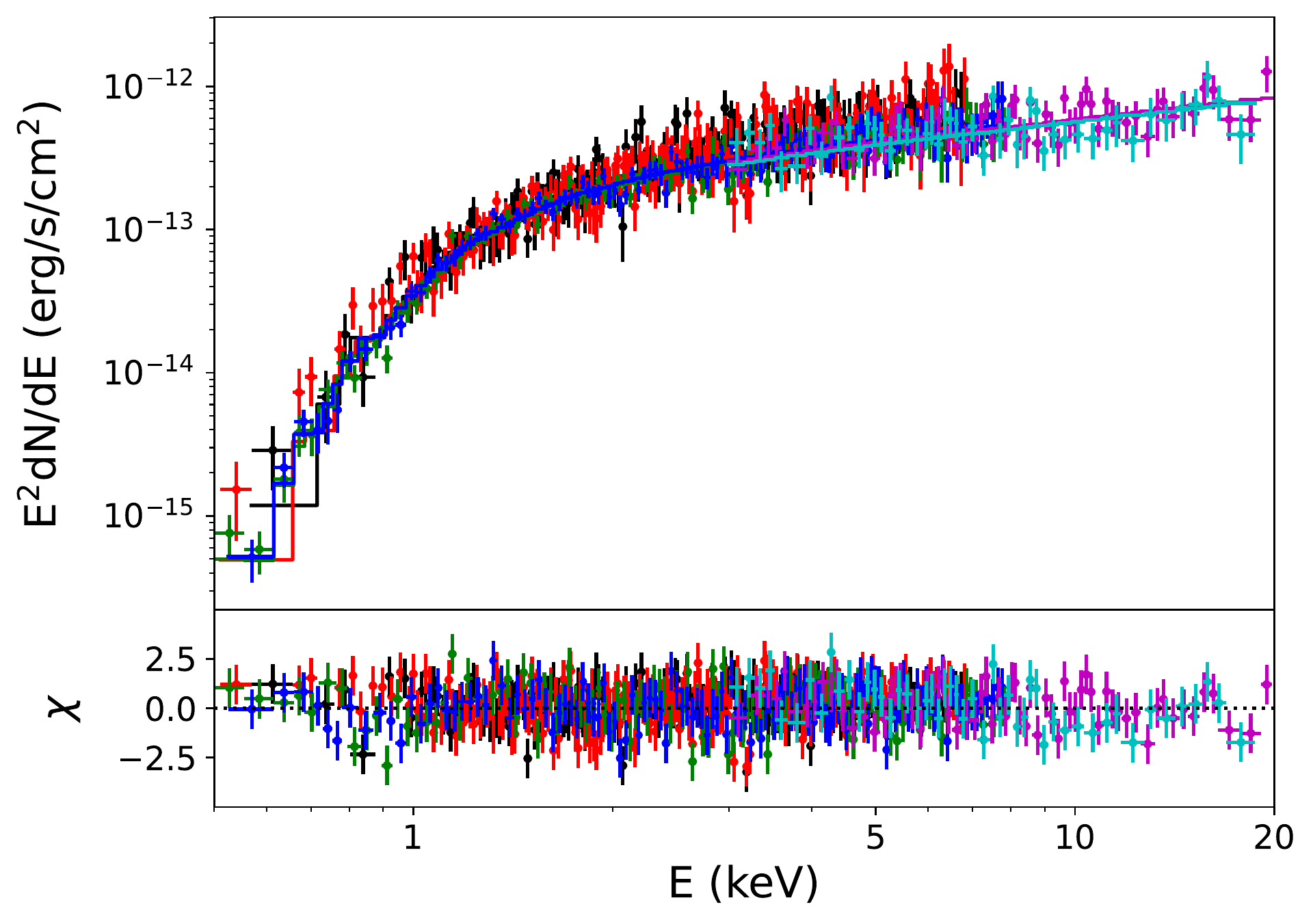}
\caption{\chandra{} ObsID 8502 (black) and 7603 (red), XMM MOS1 (green) and MOS2 (blue), \nustar{} FPMA (magenta) and FPMB (cyan) spectra of the inner nebula. The background is dominant outside of $0.5-7$ keV for \chandra{}, $0.5-8$ keV for XMM, and $3-20$ keV for \nustar{}. The source spectra were extracted from an annulus region with radii $2-20\arcsec$ for \chandra{}, a circular region with radius $40\arcsec$ for XMM, and a circular region with radius $1\arcmin$ for \nustar{}, all centered at \psr{}. The background spectra were extracted from a $2\arcmin\times2\arcmin$ box in a source-free region. For XMM and \nustar{}, the pulsar spectra were subtracted. The best-fit models are displayed as solid lines in both plots.
\label{fig:x-ray_inner_spec}}
\end{figure}

We first modeled the \chandra{} spectra of the inner nebula using an absorbed power law (\texttt{const*tbabs*pow}) to find the best-fit $N_H=(0.76\pm0.06)\times10^{22}$ cm$^{-2}$. Using this $N_H$, the best-fit model for the pulsar was found ($kT=0.13\pm0.01$ keV, $\Gamma=1.63\pm0.17$, see \S \ref{subsubsec:pulsar_spec}). This pulsar component was included and held fixed in the model for the XMM and \nustar{} spectra of the inner nebula. The $N_H$ for both instruments were held fixed to the value found from the \chandra{}-only fit. The best-fit $\Gamma$ for \chandra{} and XMM are in good agreement ($1.25\pm0.06$ and $1.35\pm0.03$, respectively), while the \nustar{} spectra give much softer $\Gamma=1.73\pm0.07$. We jointly fitted the \chandra{}, XMM, and \nustar{} spectra to constrain the model for the inner nebula more tightly and to test the presence of a spectral break. The iterative fitting of the pulsar spectrum was performed in parallel (see \S \ref{subsubsec:pulsar_spec}). An absorbed power law model with $N_H=(0.96\pm0.04)\times10^{22}$ cm$^{-2}$ and $\Gamma=1.49\pm0.03$ explains the spectra well ($\chi^2/d.o.f=710/705$). Adding a break to the power law does improve the fit (F test probability = 0.002); however, the break energy $E_{break}=6.02^{+0.75}_{-1.19}$ keV near the borderline between the \chandra{} and XMM vs. \nustar{} energy ranges is suspect. We concluded that the hint of spectral break might originate from the imperfect cross-calibration between the different instruments. The unabsorbed flux of the inner nebula is $F_{3-10}=(5.31\pm0.21)\times10^{-13}$ erg/s/cm$^2$. The best-fit $\Gamma$ for the inner nebula from the joint fit is comparable with \citet{Etten08} ($1.45\pm0.09$).

\begin{figure}[t!]
\centering
\includegraphics[width=0.45\textwidth]{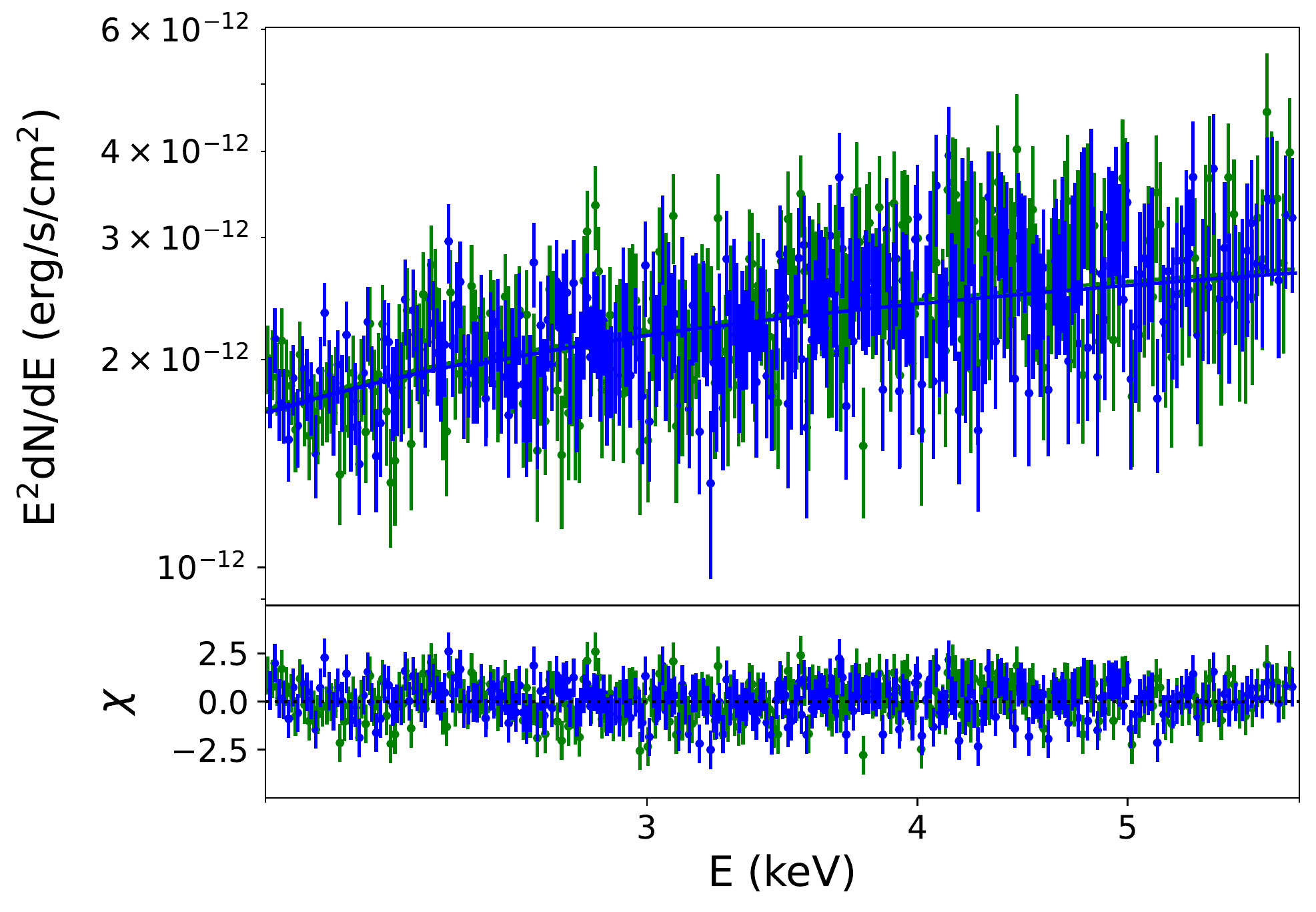}
\caption{XMM MOS1 and MOS2 (green and blue, respectively) spectra of the outer nebula. The source spectra were extracted from the dashed polygon in Figure \ref{fig:xmm_image}. The Line and continuum backgrounds are dominant below 2 keV and above 6 keV, respectively. The best-fit models are displayed as solid lines.
\label{fig:x-ray_outer_spec}}
\end{figure}

\begin{table*}[t!]
\renewcommand{\thetable}{\arabic{table}}
\centering
\caption{Summary of X-ray spectral analysis results}
\label{table:xray_spec}
\begin{tabular*}{\linewidth}{@{\extracolsep{\fill}}cccccccc}
\tablewidth{0pt}
\hline
\hline
Region & Instrument$^{\dagger}$ & Energy & $N_H$ & $kT$ & $\Gamma$ & $F_{3-10}$ & $\chi^2/d.o.f$ \\
  & & (keV) & ($10^{22}$ cm$^{-2}$) & (keV) & & ($10^{-13}$ erg/s/cm$^2$) & \\
\hline
Pulsar & C & $0.5-7$ & $0.96^{\ddagger}$ & $0.11\pm0.01$ & $1.77\pm0.17$ & $1.20^{+0.18}_{-0.17}$ & 128/151 \\
\hline
 & C & $0.5-7$ & $0.76\pm0.06$ & $-$ & $1.26\pm0.06$ & $6.80_{-0.31}^{+0.32}$ & 345/349 \\
Inner & X & $0.5-8$ & $0.76^{\ddagger}$ & $-$ & $1.35\pm0.03$ & $5.58_{-0.17}^{+0.18}$ & 246/240 \\
nebula & N & $3-20$ & $0.76^{\ddagger}$ & $-$ & $1.73\pm0.07$ & $5.50\pm0.22$ & 106/114 \\
 & C+X+N & $0.5-20$ & $0.96\pm0.04$ & $-$ & $1.49\pm0.03$ & $5.31\pm0.21$ & 710/705 \\
\hline
Outer nebula & X & $2-6$ & $0.96^{\ddagger}$ & $-$ & $1.82\pm0.03$ & $32.53\pm0.69$ & 509/541 \\
\hline
\multicolumn{8}{l}{$\dagger$ C = \chandra{}, X = XMM, N = \nustar{}. $\ddagger$ held fixed.} \\
\end{tabular*}
\end{table*}

\subsubsection{Outer nebula spectrum} \label{subsubsec:outer_spec}

We used the XMM data to analyze the outer nebula spectrum. The cleaned event files and the Swiss cheese masks (see \S \ref{subsubsec:xmm_image}) were processed with the \texttt{mos-spectra} task to extract the source spectra from the dashed polygon in Figure \ref{fig:xmm_image}. We followed the procedures described in the manual for the use of \texttt{XMM-ESAS}\footnote{\url{https://heasarc.gsfc.nasa.gov/FTP/xmm/software/xmm-esas/xmm-esas-v13.pdf}} to carefully estimate the background in such a large source region ($\sim10\arcmin$). First, the QPB spectra were generated for the same region using the \texttt{mos\_back} task. Second, the background spectrum below 2 keV contains multiple instrumental and solar wind charge exchange (SWCX) lines. We chose $2-6$ keV for the energy range of our analysis to avoid modeling too many background components. The continuum background dominates over 6 keV. Third, we attempted to model the remaining background components on \texttt{Xspec}: SP residuals and the cosmic X-ray background (CXB). The QPB spectra were loaded as background spectra. The source spectra were binned to have at least 3$\sigma$ significance over the QPB spectra in each bin. The background from SP residuals was modeled with a power law using unitary response matrices, but none of the model parameters were constrained. Therefore we assumed that the contribution from residual SP is insignificant and excluded it from the model. The CXB was modeled with an absorbed power law (\texttt{const*tbabs*pow}). All of its model parameters were fixed to the canonical values ($\Gamma=1.41$, normalization=11.6 photons/keV/s/cm$^2$/sr at 1 keV, \citet{Luca04}) to circumvent the degeneracy between the two power-law components (CXB and the outer nebula). We used the Galactic hydrogen column density\footnote{\url{https://www.swift.ac.uk/analysis/nhtot/}} at the center of the source extraction region (1.13 cm$^{-2}$) as the $N_H$ for the CXB.

The outer nebula was modeled with an absorbed power law (\texttt{const*tbabs*pow}). The $N_H$ is not constrained in the energy range of this analysis ($2-6$ keV). We fixed the $N_H$ to the best-fit value found from the \chandra{}, XMM, and \nustar{} joint fit of the inner nebula ($0.96\times10^{22}$ cm$^{-2}$). The best-fit model with $\Gamma=1.82\pm0.03$ yields a reasonable fit ($\chi^2/d.o.f=509/541$). The unabsorbed flux of the outer nebula in $2-10$ keV is $F_{2-10}=(4.20\pm0.07)\times10^{-12}$ erg/s/cm$^2$. The best-fit $\Gamma$ is clearly harder than \citet{Mizuno17} ($\Gamma=2.10\pm0.12$), yet the flux value is comparable ($F_{2-10}\sim4.1\times10^{-12}$ erg/s/cm$^2$ for the PWN-west).

\section{Fermi analysis} \label{sec:fermi}

The gamma-ray pulsations of \psr{} was first detected by AGILE (\citet{Halpern08}) and later confirmed by \fermi{} (\citet{Abdo09}). The pulsar is registered in the most recent \fermi{}-LAT source catalog (4FGL-DR3, \citet{4FGL-DR3}) as \fgl{}. We analyzed 13-year \fermi{} data (August 2008 $-$ October 2021, MET $239557417-656813666$) to detect the GeV emission from the Dragonfly. We selected SOURCE class and FRONT+BACK type events (evclass=128, evtype=3) and used the instrument response functions (IRFs) \texttt{P8R3\_SOURCE\_V3}. The 90$^\circ$ zenith angle cut and the filter expression \texttt{DATA\_QUAL>0 \&\& LAT\_CONFIG==1} were applied. The region of interest (ROI) is a $10^\circ \times 10^\circ$ box region centered at \fgl{}.

We performed a binned likelihood analysis (spatial bin = 0.1$^\circ$, energy bin = 8 bins per decade) using \texttt{Fermipy v1.0.1} (\citet{fermipy}). The ROI model includes the 4FGL-DR2 sources (\texttt{gll\_psc\_v27.fit}, \citet{4FGL-DR2}) within a $30\times30$ box region centered on \fgl{}, the Galactic diffuse emission model (\texttt{gll\_iem\_v07.fits}), and the isotropic emission model (\texttt{iso\_P8R3\_SOURCE\_V3\_v1.txt})\footnote{\url{https://fermi.gsfc.nasa.gov/ssc/data/access/lat/BackgroundModels.html}}. We used the \texttt{optimize()} and \texttt{fit()} methods in \texttt{Fermipy} to optimize the model in 100 MeV$-$300 GeV. For the \texttt{fit()} method, the parameters of bright nearby sources (within 5$^\circ$ of \fgl{} and TS (test statistics) $>25$) were left free. After fitting, the residual map was visually inspected, and a standard normal distribution was fitted to the residual significance histogram to ensure that the residuals are statistical fluctuations. 

\fgl{} is modeled with \texttt{PLSuperExp- Cutoff2}\footnote{\url{https://fermi.gsfc.nasa.gov/ssc/data/analysis/scitools/source_models.html}}. The best-fit parameters of \fgl{} agreed with those of 4FGL-DR2 within 1$\sigma$ error. Since the emission from \fgl{} cuts off in the 10$-$30 GeV range, we created a $\sqrt{\textrm{TS}}$ map of the ROI above 30 GeV to avoid contamination by the pulsar and investigate any possible diffuse emission from the PWN. We did not find any excess in the vicinity of \fgl{} that can be attributed to the emission from the PWN. This result confirms the non-detection of the GeV PWN in the vicinity of \psr{} from the previous studies of the off-pulse data (\citet{Abdo09}).

GeV gamma rays are IC upscattered photons off the low energy electrons that emit synchrotron radiation in radio $-$ infrared. Given the large ($>20\arcmin$) size of the radio nebula, a putative GeV nebula may be largely extended and too faint to be significantly detected over the background. The large ($\sim1^{\circ}$) size of the IC nebula in the VHE range (\ehwc{}) may also indicate a largely extended GeV nebula. \citet{Acero13} calculated upper limits for a GeV PWN of \psr{} assuming a size of \mgro{}. \citet{Mauro21} used an ICS template with the best-fit diffusion coefficient for \ehwc{} to place GeV upper limits of the PWN. Both works resulted in GeV upper limits similar to the flux of \ehwc{}.

\begin{figure}[t!]
\includegraphics[width=0.45\textwidth]{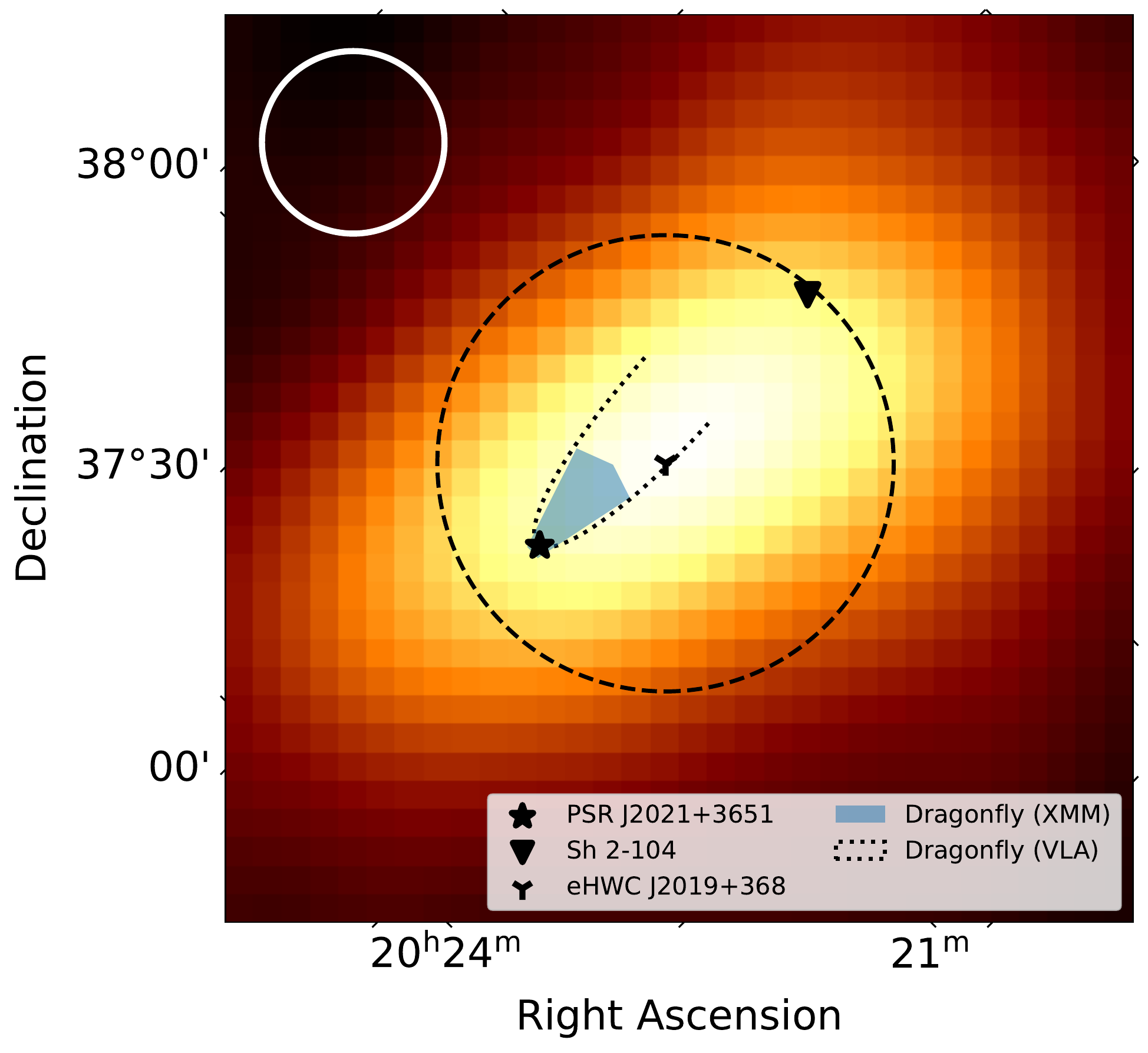}
\caption{HAWC significance map ($2^{\circ}\times2^{\circ}$) centered at the centroid of \ehwc{}. The \ehwc{} flux extraction region is marked with a dashed circle (radius $0.5^{\circ}$). \psr{} and Sh 2-104 are marked with a star and a triangle, respectively. The shaded region is the XMM spectrum extraction region of the outer nebula. The dotted line shows the extent of the outer nebula seen by VLA. The angular resolution of HAWC varies depending on the energy and zenith angle. The approximate size of the 68\% containment region at 56 TeV (radius $0.2^{\circ}$) is marked with a solid white circle at the top left corner.} The HAWC image was obtained from the 3HWC survey public data (\url{https://data.hawc-observatory.org/datasets/3hwc-survey/fitsmaps.php}), and the VLA nebular extent was estimated from \citet{Roberts08}.
\label{fig:mw_image}
\end{figure}

\section{MW SED modeling} \label{sec:sed}

Figure \ref{fig:mw_image} shows the MW counterparts of the Dragonfly overlaid on the HAWC significance map. \psr{} is located at the Eastern edge of the extended TeV source \ehwc{}. Its PWN, the Dragonfly, extends toward the centroid of \ehwc{}. We model the SED of the Dragonfly using these MW data to investigate the Dragonfly as a leptonic PeVatron.

While our \nustar{} observation allowed an in-depth study of the inner nebula, the faint emission from the outer nebula was not detected due to the limited sensitivity. Instead, we used our XMM analysis result of the outer nebula presented in \S \ref{subsubsec:outer_spec}. The radio spectrum and the GeV upper limits were taken from \citet{Roberts08} and \citet{Mauro21} (``IEM-\textit{4FGL}"), respectively. In the TeV band, three independent flux measurements by VERITAS (\verone{}), HAWC (\ehwc{}), and LHAASO (\lhaaso{}) are available. \citet{Cao21a} did not provide detailed spectral information of \lhaaso{} except for its flux at 100 TeV. \citet{Abeysekara18} and \citet{Abeyesekara20} provide the spectrum of \verone{} and \ehwc{}, respectively, over three decades of energy. While \verone{} and \ehwc{} exhibit similar source size, the flux of \verone{} reported in \citet{Abeysekara18} was extracted from a region smaller than the source size, yielding a $2-3$ times lower flux than that of \ehwc{} in the overlapping energy range (see also \citet{Albert21}). Therefore, we used the spectrum of \ehwc{} from \citet{Abeyesekara20} for SED modeling in this work. 

The distance estimates of \psr{} vary widely depending on different distance measures. \citet{Roberts02} used the dispersion measure (DM) $\sim 371$ pc cm$^{-3}$ to put the pulsar at $\geq$ 10 kpc on the outer edge of the Galaxy, although they left a possibility of a nearer distance in case of a contribution from excess gas in the Cygnus region. \citet{Etten08} suggested 3$-$4 kpc based on various arguments, such as the X-ray spectral fit to a neutron star atmosphere model and the gamma-ray efficiency of the pulsar. \citet{Abdo09} estimated a distance $\geq$ 4 kpc based on the pulsar rotation measure (RM). \citet{Kirichenko15} suggested $1.8^{+1.7}_{-1.4}$ kpc using the interstellar extinction and distance relation. The 1.8 kpc distance was adopted by \citet{Mizuno17}, \citet{Fang20}, and \citet{Albert21} for their MW SED modeling.

We start with preliminary modeling of the MW SED using \texttt{Namia} (\citet{Naima}), a generic model for non-thermal radiation from relativistic particles. We do not make assumptions about the distance or evolutionary history of the PWN at this stage. Our purpose for this preliminary modeling is to provide a basic understanding of the current status of the PWN and initial estimation of the input parameters for a more sophisticated model, namely the dynamical model (\citet{Gelfand09}). Then we move on to MW SED modeling using the dynamical model to acquire insight into the dynamical evolution of the Dragonfly over its lifetime while the interactions between the PWN, SNR, and ISM are accounted for.

\subsection{Naima} \label{subsec:naima}

\texttt{Naima} allows us to characterize the current particle population using a minimal number of parameters without introducing any physical assumptions on the evolutionary history of the system.

\begin{figure}[t!]
\includegraphics[width=0.45\textwidth]{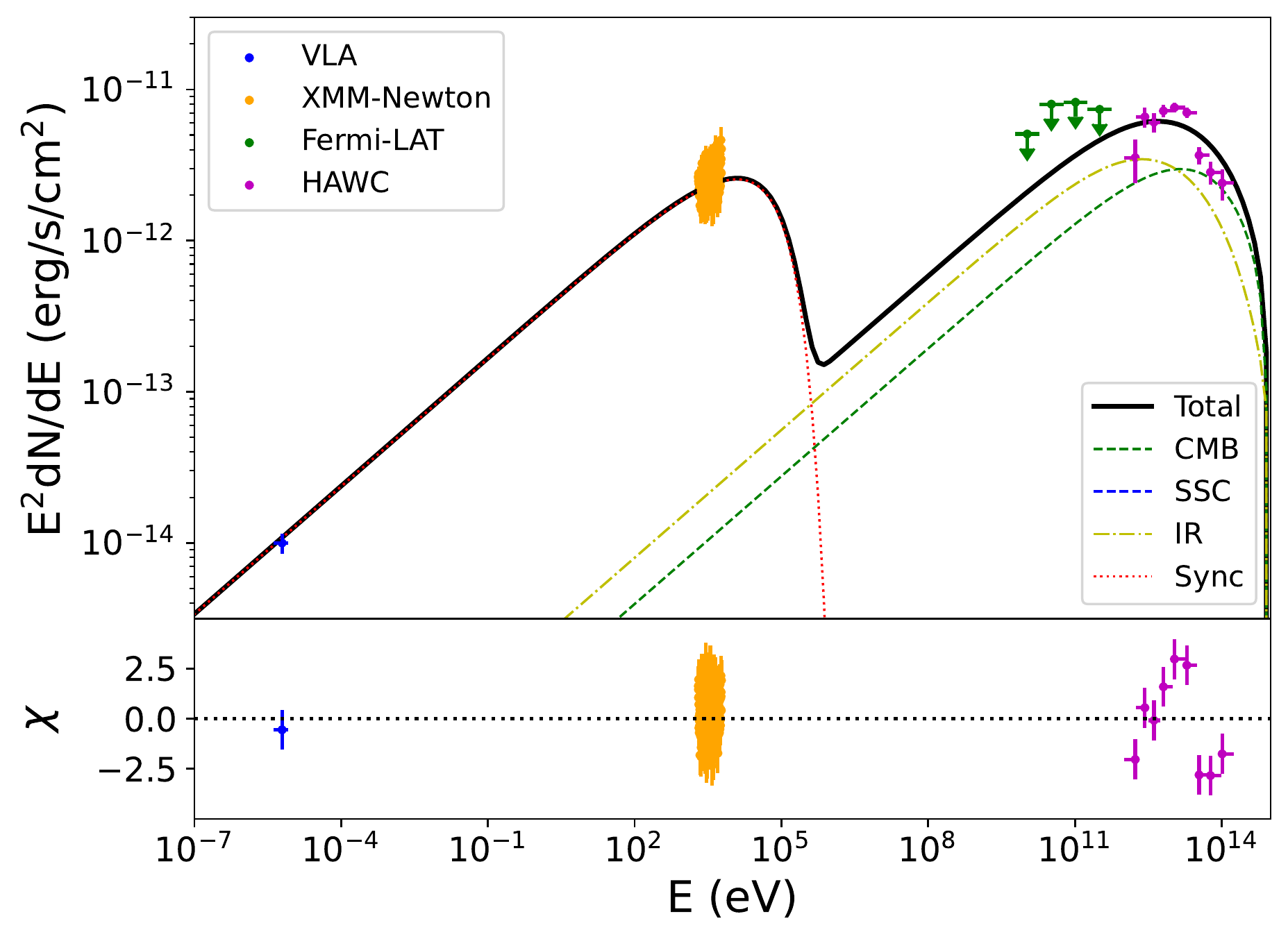}
\caption{The best-fit SED model from Naima. The model parameters are given in Table \ref{table:naima}. Synchrotron flux (Sync), inverse Compton flux from the CMB (CMB), infrared emission from the interstellar dust grains (IR), synchrotron self-Compton component (SSC), and total flux are plotted. The SSC flux level is very low and located below the lower bound of the y-axis. Radio, X-ray, and TeV flux data points, and GeV upper limits are overlaid. The residuals are plotted in terms of significance ((data$-$model) / (1$\sigma$ uncertainty of data)). 
\label{fig:naima}}
\end{figure}

For a leptonic particle accelerator, synchrotron (SC) emission and emission via inverse Compton (IC) scattering off the input seed photon fields (cosmic microwave background (CMB) and interstellar dust emission (infrared, or IR)) are calculated based on a particle distribution model. We vary the model parameters of a single particle distribution so that the SC and IC spectra are consistent with the observed flux  in radio, X-ray, and TeV gamma-ray bands. The minimum particle energy ($E_{min}$) and the reference energy ($E_{0}$) were fixed to 1 MeV and 1 TeV, respectively. The best-fit parameters are summarized in Table \ref{table:naima}, and the best-fit SED model is plotted with the MW data and residuals in Figure \ref{fig:naima}.

\begin{table}[t!]
\renewcommand{\thetable}{\arabic{table}}
\centering
\caption{Best-fit SED model parameters using Naima}
\label{table:naima}
\begin{tabular}{cc}
\tablewidth{0pt}
\hline
\hline
$\alpha$ & 2.4 \\
$E_{cut}$ & 0.9 PeV\\
Magnetic field & 1.6 $\mu$G \\
IR temperature & 26 K \\
IR energy density & 1.0 eV/cm$^3$\\
Total particle energy$^{\dagger}$ & $6.1d_{3.5}^2\times10^{49}$ erg \\
\hline
\multicolumn{2}{l}{$\dagger$ $d_{3.5}$ is the distance to the Dragonfly scaled to the} \\
\multicolumn{2}{l}{nominal distance of 3.5 kpc.} \\
\end{tabular}
\end{table}

The TeV spectrum exhibits a smooth cutoff after 20 TeV. This cutoff is better explained by an exponential cutoff power law distribution of particles, $dN/dE=A(E/E_0)^{-\alpha}e^{-E/E_{cut}}$, than a simple power law with a sharp cutoff at $E_{max}$. The best-fit particle index is $\alpha=2.4$. Adding an IR field gives a better fit than the CMB-only model, although the IR field energy density tends to grow indefinitely to an unphysical value. Therefore we fixed the IR field energy density to the average cosmic ray energy density (1 eV/cm$^3$, \citet{Cummings16}). This yields the best-fit IR field temperature T = 26 K, magnetic field B = 1.6 $\mu$G, and cutoff energy $E_{cut}=0.9$ PeV. This SED model shows a good fit to the MW data as seen in the residuals plotted in Figure \ref{fig:naima}; however, the narrow IC peak resolved by HAWC is difficult to explain with physically reasonable model parameters. Such a narrow peak originates from the flux point in the lowest energy bin, the range in which air imaging Cherenkov telescopes (IACTs), such as VERITAS, are more sensitive. A deeper VERITAS observation and more accurate flux measurement of the region have been proposed for our future work to resolve the IC spectrum of the Dragonfly better.

The best-fit cutoff energy $E_{cut}=0.9$ PeV strongly suggests that the Dragonfly is likely a PeVatorn. The cutoff energy is greater than 0.3 PeV by \citet{Albert21} and 0.4 PeV by \citet{Fang20} mainly due to the difference in the X-ray spectra used in each work (see \S \ref{sec:sed}). The magnetic field inside the PWN is at the level of interstellar magnetic field ($\sim$ 3 $\mu$G, \citet{Jansson12}, as one may infer from the low X-ray to gamma-ray luminosity ratio.

The IR field temperature is higher than the IR emission from cold dust grains ($\sim 15$ K). A possible source of this warm dust emission is an H II region Sh2-104 (marked with an inverted triangle in Figure \ref{fig:mw_image}). Located at $4 \pm 0.5$ kpc from the Earth, Sh2-104 is visible in radio through X-ray (\citet{Deharveng03}, \citet{Rodon10}, \citet{Xu17}, and \citet{Gotthelf16}). Sh 2-104 hosts an ultra-compact H II (UCHII) region on the eastern periphery of its dense molecular shell. As strong candidates for active star formation (\citet{Deharveng03} and \citet{Xu17}), Sh2-104 and the UCHII region each contain a stellar cluster, MASS J20174184+3645264 and IRAS 20160+3636, respectively, ionizing the regions (\citet{Deharveng03} and \citet{Paredes09}). \citet{Rodon10} used Herschel observations to estimate the dust temperature in Sh2-104 to be $20-30$ K on the exterior. Our best-fit IR field temperature (26 K) lies in this temperature range. Since the lower bound of the distance to Sh2-104 (\citet{Deharveng03}) and the upper bound of the distance to \psr{} \citet{Kirichenko15} coincide (3.5 kpc), we adopt 3.5 kpc as a nominal distance to the Dragonfly hereafter and scale relevant parameters to this distance whenever possible. 

\subsection{Dynamical model} \label{subsec:dynamic}

Taking the result from \texttt{Naima} as a starting point, we fit the dynamical model to the MW data of the Dragonfly. The dynamical model is a time-evolutionary model for a composite system of a PWN and its host SNR. The model assumes a spherical single-zone system whose SNR is in a free-expansion or Sedov-Taylor phase. The model evolves a PWN and its SNR from their birth to the true age of the system, calculating the interaction between them and with the surrounding interstellar medium (ISM). The model output includes a pulsar wind particle distribution, its synchrotron and inverse Compton emission spectrum, and the dynamics of a system (e.g., a radius of a PWN, a radius of an SNR forward and reverse shock, and a magnetic field inside a PWN) at each evolutionary phase.

\begin{table}[t!]
\renewcommand{\thetable}{\arabic{table}}
\centering
\caption{SED model parameters using the dynamical model}
\label{table:dynamical}
\begin{tabular}{ccc}
\tablewidth{0pt}
\hline
\hline
\multicolumn{2}{c}{Source distance} & 3.5 kpc \\
\hline
\multirow{8}{*}{Input}
& $E_{SN}$ & $1.0\times10^{51}$ erg \\
& $M_{ej}$ & $7.2 M_{\odot}$ \\
& $n_{ISM}$ & 0.03 cm$^{-3}$ \\
& Braking index $p$ & 2.5 \\
& Age & 16 kyr \\
& $\eta_{B}$ & 0.008 \\
& $E_{max}$ & 1.4 PeV \\
& Particle index $\alpha$ & 2.4 \\
& IR field temperature & 9.9 K \\
& IR field energy density & 1.4 eV cm$^{-3}$ \\
\hline
\multirow{4}{*}{Output}
& $t_{RS}$ & 12 kyr \\
& $R_{PWN}$ & 9.5 pc \\
& Magnetic field & 2.7 $\mu$G \\
& Total particle energy & $3.9\times10^{48}$ erg \\
\hline
\end{tabular}
\end{table}

The dynamical model evolves the particle distribution inside a PWN via three mechanisms: continuous particle injection of a fraction of the pulsar spin-down luminosity, adiabatic energy loss due to the expansion of the PWN, and radiative energy loss due to synchrotron and inverse Compton emission. The spin-down luminosity $\dot{E}$ of a PWN at its age $t$ is formulated as
\begin{equation} \label{eq:edot}
    \dot{E}(t)=\dot{E}_0\left(1+\frac{t}{\tau_{sd}}\right)^{-\frac{p+1}{p-1}},
\end{equation}
where $\dot{E}_0$ is the initial spin-down luminosity, $p$ is the pulsar braking index, and $\tau_{sd}$ is the characteristic pulsar spin-down timescale. $\tau_{sd}$ is related to a pulsar's characteristic age $\tau$ and true age $t_{age}$ as
\begin{equation} \label{eq:tage}
    t_{age}=\frac{2\tau}{p-1}-\tau_{sd}.
\end{equation}

A fraction of the spin-down luminosity $\eta_B\dot{E}$ is injected into the PWN as magnetic fields, while the rest of the spin-down luminosity, $(1-\eta_B)\dot{E}$, is injected as particles (electrons and positrons). The particle injection spectrum is defined within the energy range between $E_{min}$ (minimum energy of the injected particle, fixed to 0.1 GeV in this work) and $E_{max}$ (maximum energy of the injected particle). The magnetic field is assumed to be homogeneous throughout the volume of the PWN, and thus it decreases as the PWN expands and increases as the PWN is crushed by the collision with the SNR reverse shock. The radiative loss changes accordingly $-$ synchrotron emission is much stronger than inverse Compton emission in the early stage or post-collision era of the PWN, whereas inverse Compton flux becomes comparable with synchrotron flux as the PWN ages. Adiabatic loss is most severe when a PWN freely expands against only ram pressure from the unshocked SN ejecta in the free-expansion phase. Once the PWN collides with the SNR reverse shock and starts encountering the pressure from shocked ejecta, the expansion of the PWN slows down until the compression starts, during which the PWN undergoes adiabatic heating.

\begin{figure*}[t!]
\gridline{\fig{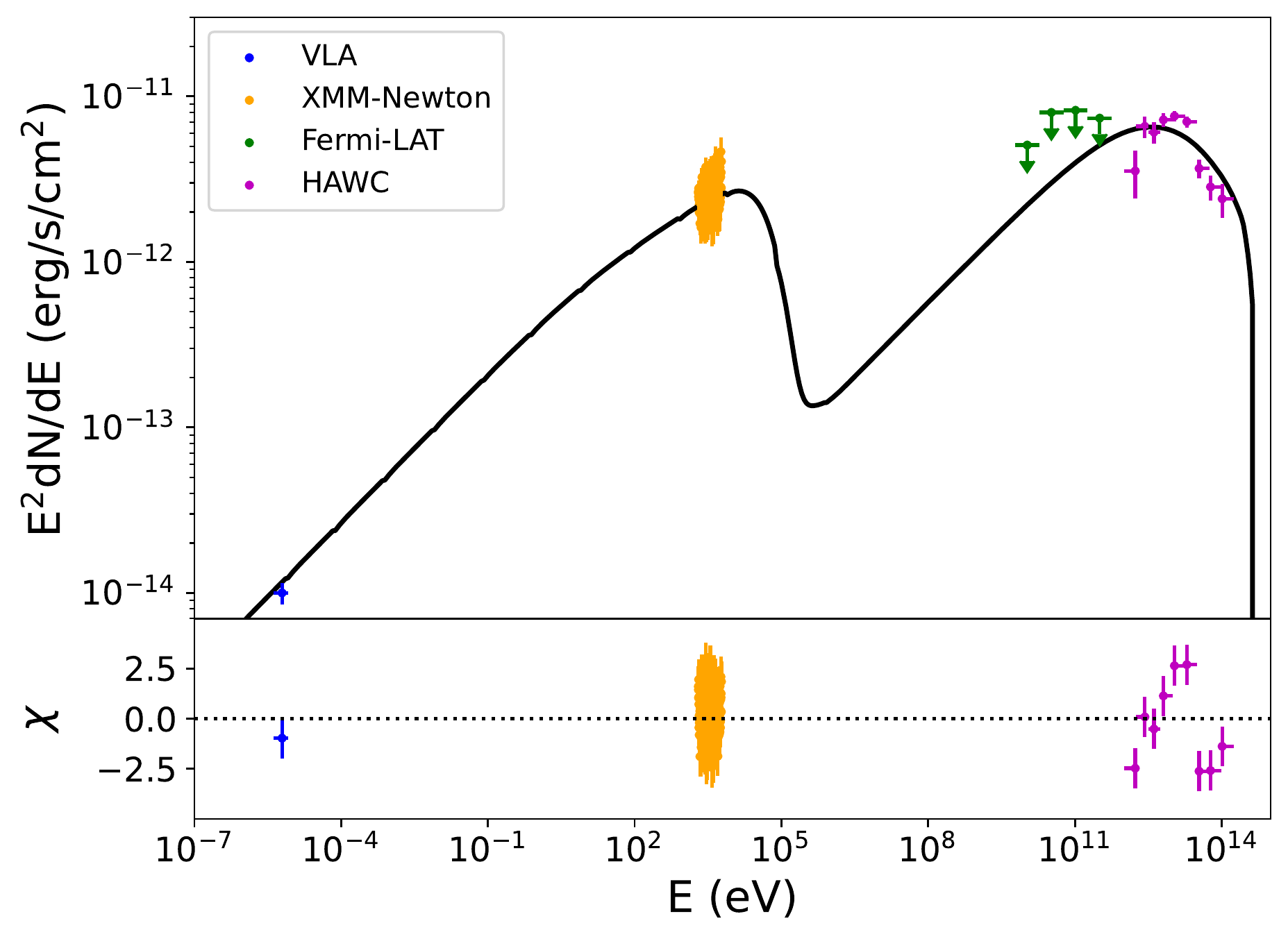}{0.43\textwidth}{(a)}
\fig{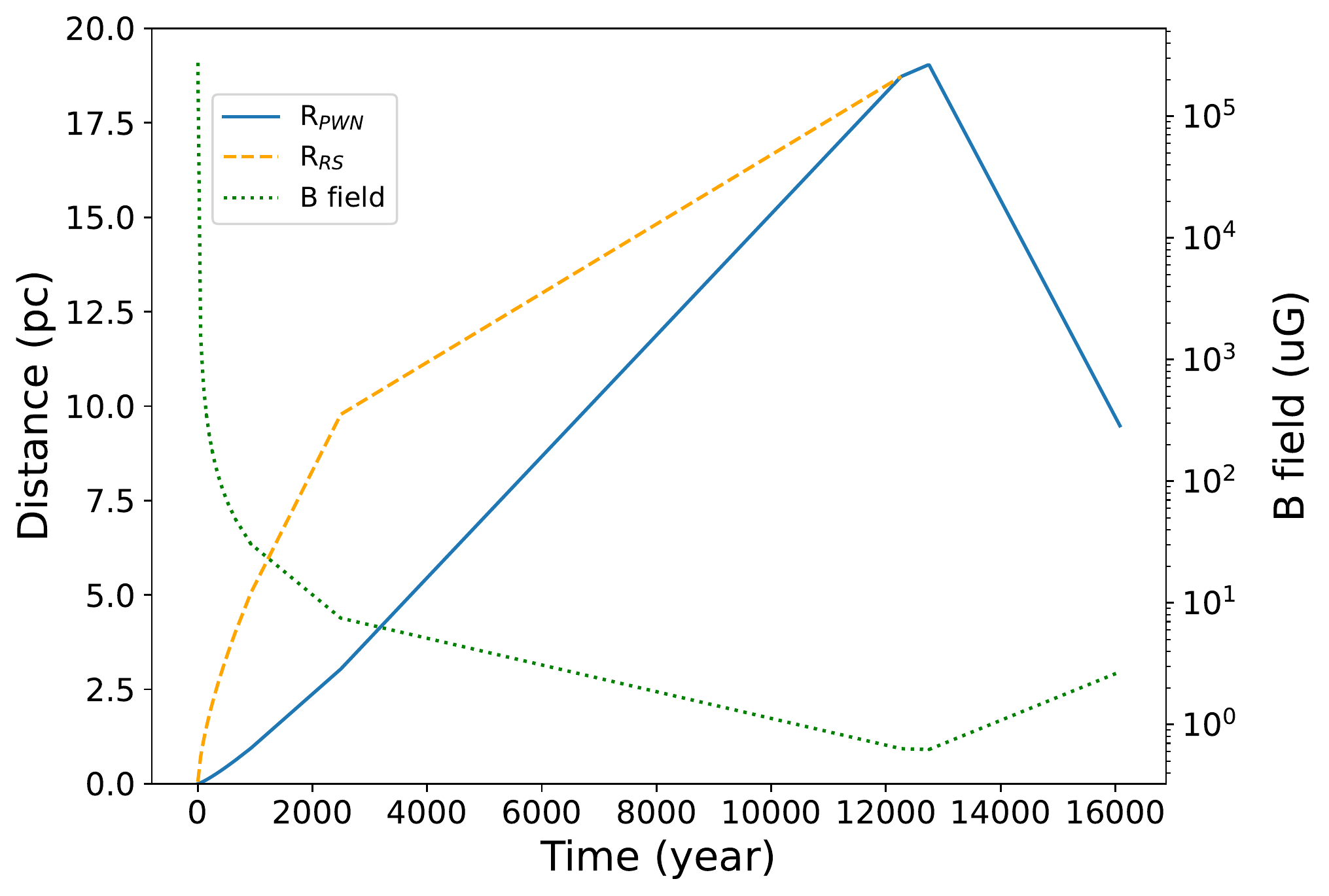}{0.47\textwidth}{(b)}}
\gridline{\fig{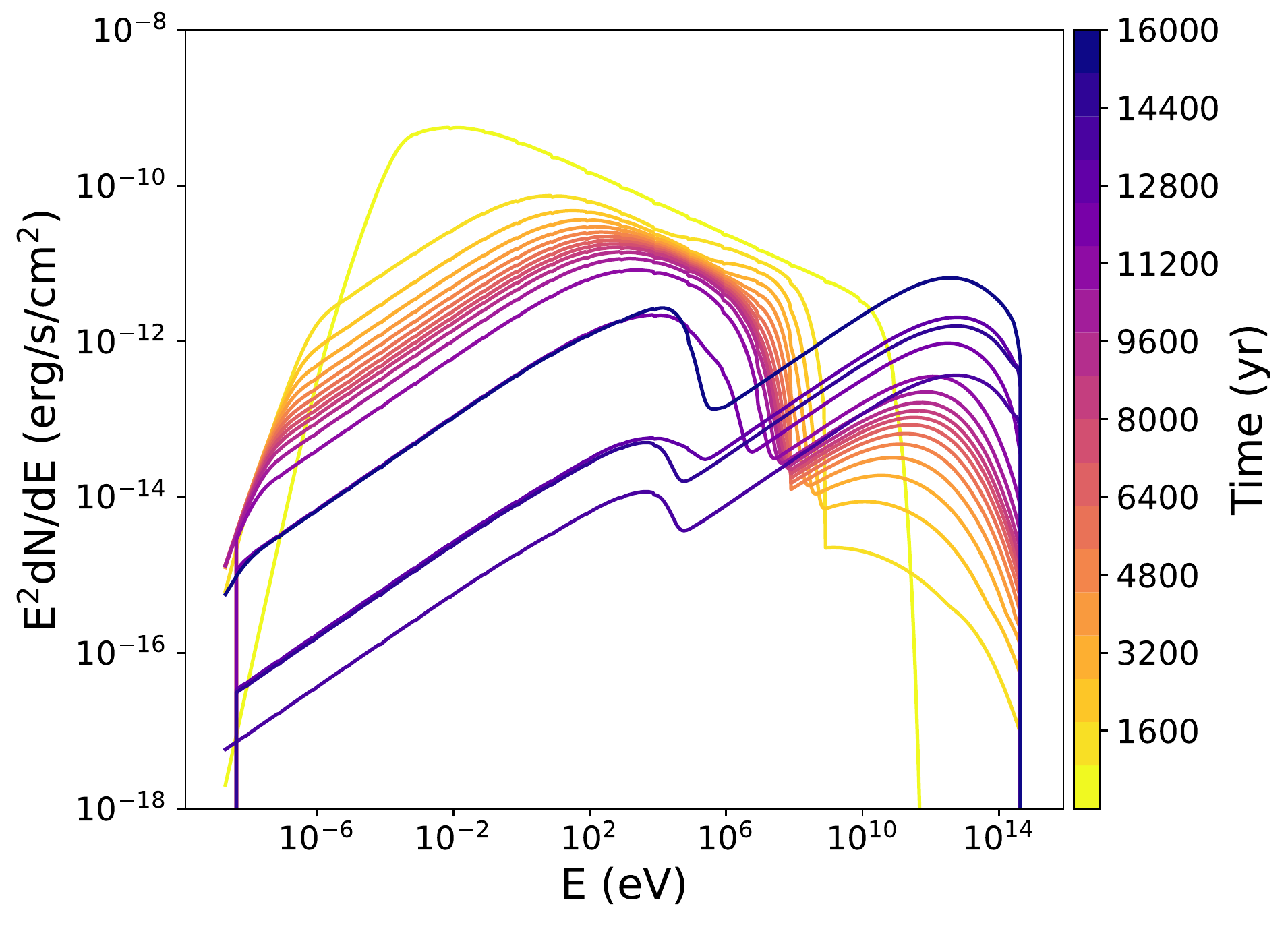}{0.45\textwidth}{(c)}
\fig{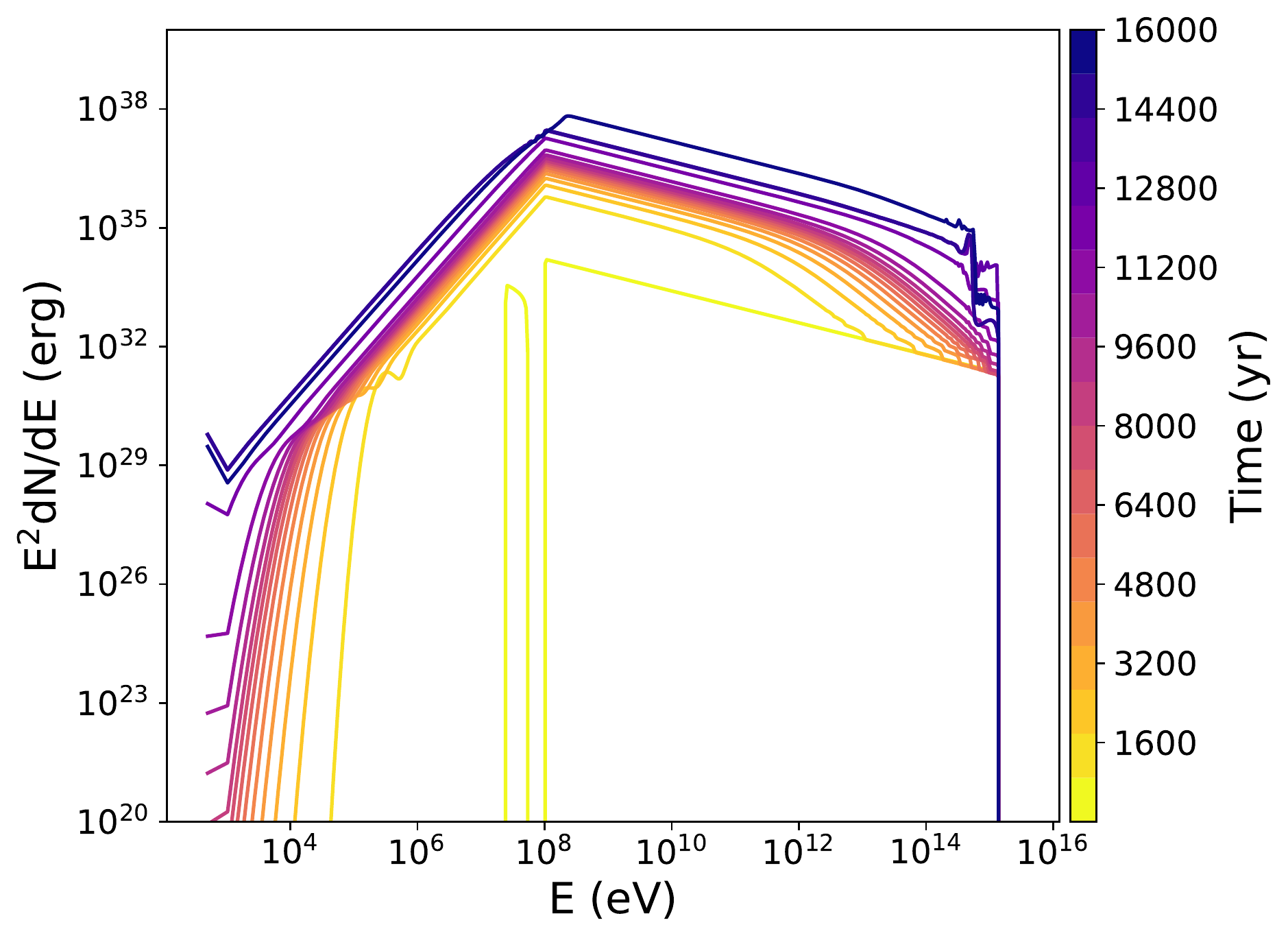}{0.45\textwidth}{(d)}}
\caption{(a) The SED model for the Dragonfly from the dynamical model. The model parameters are given in Table \ref{table:dynamical}. Radio, X-ray, and TeV flux data points are overlaid. The residuals are plotted in terms of significance ((data$-$model)/(1$\sigma$ uncertainty of data)). (b) Time evolution of the PWN radius (R$_{pwn}$, blue solid line), SNR reverse shock radius (R$_{RS}$, orange dashed line), and magnetic field inside the PWN (B field, green dotted line) over the true age from our model (16 kyr). The PWN collided with the SNR reverse shock at $t_{RS}=12$ kyr. (c)(d) Time evolution of the radiation (c) and particle (d) spectrum of the Dragonfly.}
\label{fig:dynamical_sed}
\end{figure*}

The dynamical evolution of a system is calculated based on input parameters related to a supernova (SN), SNR, and surroundings, such as SN explosion energy ($E_{SN}$), ejecta mass inside an SNR ($M_{ej}$), and ISM density just outside an SNR forward shock ($n_{ISM}$). These parameters determine the pressure just outside a PWN and at the location of a reverse shock. A PWN size changes such that the pressure from the ejecta is in balance with the pressure inside a PWN, which comprises magnetic pressure and pressure from particles as a relativistic ideal gas. When a reverse shock reaches a PWN, the pressure experienced by a PWN increases dramatically. This leads to a rapid decrease in the size of a PWN and, consequently, a sharp increase in the magnetic field inside a PWN. Synchrotron loss is extreme at this point. Once a PWN is highly compressed, and its pressure exceeds the ejecta pressure, the PWN starts re-expanding, and hence the magnetic field inside the PWN starts decreasing. As a PWN ages, its size grows as large as several parsecs or more, and its magnetic field becomes as low as a few $\mu$G. The resulting spectrum yields a similar level of SC and IC flux, a typical spectrum observed in middle-aged PWNe. 

\begin{table*}[t!]
\renewcommand{\thetable}{\arabic{table}}
\centering
\caption{Comparison between PeVatron pulsar wind nebulae}
\label{table:pevatron}
\begin{tabular*}{\linewidth}{@{\extracolsep{\fill}}ccccc}
\tablewidth{0pt}
\hline
\hline
\multicolumn{2}{c}{ } & G75.23+0.12 (``Dragonfly") & G18.5-0.4 (``Eel")$^a$ & G106.65+2.96 (``Boomerang")$^b$ \\
 \hline
\multicolumn{2}{c}{\multirow{2}{*}{TeV counterpart}}
& \ehwc{} & HAWC J1826-128 & HAWC J2227+610 \\
\multicolumn{2}{c}{ } & \lhaaso{} & LHAASO J1825-1326 & LHAASO J2226+6057 \\
\hline
\multirow{4}{*}{Pulsar}
& Name & \psr{} & PSR J1825-1256 & PSR J2220+6114 \\
& $\tau$ (kyr) & 17 & 14 & 10 \\
& $\dot{E}$ ($10^{36}$ erg/s) & 3.4 & 3.6 & 22 \\
& Distance$^c$ (kpc) & 0.4 $-$ 12 (3.5) & 3.5 & 0.8 $-$ 7.5 (7.5) \\
\hline
\multirow{4}{*}{PWN}
& $E_{max}$ (PeV) & 1.4 & 4.6 & 3.3 \\
& Magnetic field ($\mu$G) & 2.7 & 0.6 & 2.2 \\
& True age (kyr) & 16 & 4.6 & 3.3 \\
& $t_{RS}$ (kyr) & 12 & $-$ & 1.5 \\
\hline
\multicolumn{5}{l}{$^a$\citet{Burgess22} and references therein. $^b$Pope et al. (submitted to ApJ) and references therein.} \\
\multicolumn{5}{l}{$^c$For the sources with a wide range of distance estimates, the distance used for SED modeling is given in parentheses.} \\
\end{tabular*}
\end{table*}

We aimed to find a set of model parameters that reproduces not only the MW spectrum but also the observed size of the Dragonfly. The Dragonfly displays a highly asymmetric morphology, although the dynamical model assumes a spherical system. Given this limitation of the model, we focus on characterizing the evolution of the PWN properties averaged over the entire system rather than their spatial dependency. We approximated a nominal radius of the PWN to be 10$\arcmin$ so that its spherical volume roughly matches the physical volume of the outer nebula. The radio nebula size $\sim 20\arcmin$ was used for this calculation $-$ the lowest energy particles seen in the radio band would have a much longer lifetime than its synchrotron cooling time and thus best reflect the true extent of the PWN. The angular size of 10$\arcmin$ is equivalent to $R_{PWN}=10$ pc at the nominal distance $d=3.5$ kpc. 

The input and output parameters of the dynamical model are listed in Table \ref{table:dynamical}. Figure \ref{fig:dynamical_sed} shows the SED model plotted with the MW data and residuals, the time evolution of the dynamical parameters (magnetic field, PWN radius $R_{PWN}$, and SNR reverse shock radius $R_{RS}$), radiation and particle spectra.

The maximum particle energy $E_{max}=1.4$ PeV provides strong evidence that the Dragonfly is a PeVatron PWN. The true age of 16 kyr was found to be slightly younger than its characteristic age $\tau=17$ kyr. The true age found by our model is much older than 7 kyr found by \citet{Albert21}. This difference can be attributed most likely to the assumed source distance (3.5 kpc in this work, 1.8 kpc in \citet{Albert21}), as well as to the SED models and the X-ray spectra (see \S \ref{sec:sed}). The low magnetic field (2.7 $\mu$G) is consistent with that from \texttt{Naima}, \citet{Mizuno17}, \citet{Fang20}, and \citet{Albert21}. The low magnetic fraction $\eta_B=0.008$ contributes to this low magnetic field. 

Our model predicts that the PWN expanded to $\sim 20$ pc, collided with the SN reverse shock 4 kyrs ago ($t_{RS}=12$ kyr), and has been shrinking since then to reach the current size $\sim 10$ pc (Figure \ref{fig:dynamical_sed} (b)). Relatively low ISM density $n_{ISM}=0.03$ cm$^{-3}$ drove a slow reverse shock, allowing the PWN to grow large enough to reach the reverse shock even before it started heading back toward the PWN. Combined with the substantial ejecta mass $M_{ej}=7.2M_{\odot}$, this low ISM density may indicate that the host SNR of the Dragonfly evolved into the wind-blown bubble of a massive progenitor star with an extremely low density (below 0.001 cm$^{-3}$) during the first few kyrs of its lifetime (\citet{Dwarkadas05}).

The particle index $\alpha=2.4$ is consistent with \texttt{Naima}. The IR field temperature (9.9 K) falls below the range of the dust temperature in Sh2-104 (see \S \ref{subsec:naima}). Using the braking index p = 2.5 and $\tau_{sd}=6.8$ kyr of our model, the total particle energy ($3.9\times10^{48}$ erg) is 50\% of the total injected energy over the true age (16 kyr) of the Dragonfly.

\section{Discussion} \label{sec:discussion}

\subsection{PeVatron pulsar wind nebulae} \label{subsec:pevatron}

We compare three PeVatron PWNe studied in our \nustar{} observational program of energetic PWNe: G75.23+0.12 (``Dragonfly", this work), G18.5-0.4 (``Eel", \citet{Burgess22}), and G106.65+2.96 (``Boomerang", submitted to ApJ). All three PWNe were modeled with the dynamical model. Key facts and the model parameters of the three PWNe are summarized in Table \ref{table:pevatron}. The common features of the three PeVatron PWNe are the following:
\begin{enumerate}
    \item The maximum particle energy is greater than 1 PeV.
    \item The source morphology is highly asymmetric and energy-dependent. The pulsar is located on the edge of the extended radio and soft X-ray nebulae and is offset from the centroid of its TeV counterparts. \label{pev:morphology}
    \item The magnetic field strength inside the PWNe is low $<$ 3 $\mu$G. \label{pev:bfield}
    \item Compact hard X-ray nebula was detected up to 20 keV by \nustar{}. The nebular size is much smaller than the lower-energy nebula (radio and soft X-ray). \label{pev:compact}
    \item The compact hard-Xray nebular spectrum does not exhibit a spectral break or cutoff. A synchrotron burnoff is observed from the shrinkage of its size at higher energies.
    \item The dynamical model predicts that the PWN collided with the reverse shock of the host SNR ``recently" (the Dragonfly and Boomerang), or such a collision is about to happen (the Eel). \label{pev:reverse}
\end{enumerate}

\ref{pev:morphology} and \ref{pev:bfield} are known properties of bright TeV PWNe (e.g., \citet{Kargaltsev13} and \citet{Wilhelmi17}). While asymmetric morphology bears a few different possibilities (fast pulsar velocity, asymmetric SNR reverse shock $-$ PWN interaction, or a combination of both effects $-$ see \S \ref{subsec:distance}), energy dependency of morphology can be attributed to particle transport and cooling. Particles that emit synchrotron radiation in the radio band (particle energy $E_{\gamma} \sim$ 1 GeV in the interstellar magnetic field $\sim$ 3 $\mu$G) have cooling times much longer than the age of the PWN and hence transport to large distances away from the pulsar without losing much of their energies. Particles that radiate in the hard X-ray band ($E_{\gamma} \sim$ 100 TeV), on the other hand, have much shorter cooling times than the PWN age. Such particles can travel only to short distances before cooling down to lower energies, resulting in \ref{pev:compact} (see \S \ref{subsec:bfield}). Therefore, only freshly injected highly energetic particles contribute to the compact hard X-ray nebula. Relic particles, after cooling, exhibit larger extents in lower energies.

Looking at an IC spectrum, relic particles with energies $E_\gamma \sim$ 10 TeV upscatter the CMB photons to TeV energies. Such particles can be dim in the synchrotron spectrum due to a lower magnetic field farther away from the pulsar, explaining the offset of the TeV emission from the pulsar. \ref{pev:bfield} is necessary for this reason and manifests as the observed low X-ray to gamma-ray luminosity ratios. GeV-emitting particles ($E_\gamma \sim$ 10 GeV) are expected to form even fainter SC and IC nebulae due to their lower energies and larger distances traveled. No GeV nebulae were detected for the PeVatron PWNe except for the ``tail" region of the Boomerang whose emission is attributed to its parent SNR (e.g., \citet{Fang22}).

\ref{pev:reverse} provides a hint as to how particles are accelerated in PeVatorn PWNe. \citet{Ohira18} proposed using a Monte Carlo simulation that particles may be accelerated to 1 PeV during the compression of a PWN by the collision with the reverse shock of its host SNR.

Some of the above properties are in contrast to those of other TeV PWNe in our \nustar{} observational campaign, such as G313.54+0.23 (``K3" or ``Kookaburra", \citet{K3}) and G313.3+0.1 (``Rabbit", \citet{Rabbit}). These southern sources are invisible to HAWC and LHAASO, the telescopes that operate in the highest energy regime ($>$ 100 TeV). Their \nustar{} hard X-ray nebulae are extended (radius $\sim 3\arcmin$ at the source distance $\sim$ 5.6 kpc for both PWNe), and the nebular sizes do not change significantly with energy. Multi-zone SED modeling using the spatially resolved \nustar{} spectra along with MW flux data yielded the maximum particle energies below 1 PeV for both PWNe. The PWNe have bright GeV counterparts whose spectra connect smoothly to the spectra of their TeV counterparts. Future gamma-ray observatories in the southern hemisphere, the Southern Wide-field Gamma-ray Observatory (SWGO) and Cherenkov Telescope Observatory (CTAO) $-$ South, will play a crucial role in studying the true energetics of these PWNe and their relation to the MW observations.

\subsection{Distance and proper motion} \label{subsec:distance}

The distance of the Dragonfly is relevant to not only its brightness but also its physical size, and hence the proper motion of \psr{}. Like many other TeV PWNe (e.g., \citet{HESS18}), the Dragonfly is offset from the center of the nebula and its TeV counterparts. Such highly asymmetric morphology is often attributed to a fast proper motion of the pulsar. \citet{Aliu14} estimated the transverse velocity of \psr{} to be $840 d_5 t_{17}^{-1}$ km/s, where $d_5$ is the distance of the Dragonfly scaled to 5 kpc and $t_{17}$ is the age of \psr{} scaled to its characteristic age of 17 kyr, in case it was born at the end of the radio nebula. \citet{Albert21} estimated it to be $\sim$ 1,300 km/s in case \psr{} is located at 1.8 kpc and was born 7 kyrs ago at the location of HAWC J2019+368. \citet{Etten08} and \citet{Jin23} claimed the detection of a bow shock structure on the East side of the Dragonfly, yet noted a possibility of at most a mildly supersonic motion of \psr{} considering the well-preserved substructures of the inner nebula.

Given the true age of 16 kyr from our model, the Dragonfly may be too young to have escaped its host SNR and form a bow shock in the ISM (e.g., \citet{Gaensler06}). Instead, the bow shock structure with the well-defined inner nebula and the asymmetric PWN morphology could be explained by an asymmetric interaction between the PWN and the reverse shock of its host SNR due to an ambient density gradient (e.g., \citet{Temim15} and \citet{Temim17}). In this case, the orientation of the bow shock does not necessarily align with the direction of the pulsar's proper motion (\citet{Kolb17}). There is no known dense object on the East of \psr{}, and the host SNR has not been detected. This mystery could be solved by a deep and expansive radio observation that covers a large ($\sim 1^{\circ}$) region to search for the faint host SNR of \psr{}. Here, we focus on discussing the proper motion of \psr{} that may have caused the asymmetric morphology of the Dragonfly.

The angular separation between \psr{} and the centroid of \ehwc{} is $\sim 16 \arcmin$. If \psr{} was born near the centroid of \ehwc and traveled to the current location at a constant speed, the corresponding transverse velocity of the pulsar is $v_{psr}=996d_{3.5}t_{16}^{-1}$ km/s, where $t_{16}$ is the true age scaled to 16 kyr. This is above the average pulsar velocity (540 km/s, \citet{Verbunt17}), but not exceptionally high (\citet{Kargaltsev17}). In this case, measuring the pulsar proper motion of $\sim 0.06\arcsec d_{3.5}t_{16}^{-1}$/yr may not be feasible unless \psr{} is significantly closer than 3.5 kpc or significantly younger than 16 kyr. Another \chandra{} observation of \psr{}, nearly 20 years after the last observation, to detect the pulsar motion could provide insight into the source distance and age; however, their degeneracy will still need to be disentangled. Our future work will combine new radio (VLA) and X-ray (\chandra{}) observation with an energy-dependent morphology study using VERITAS and \fermi{}-LAT to place tight constraints on the source distance and evolutionary history of the Dragonfly.

\subsection{Magnetic field} \label{subsec:bfield}

For particles with a synchrotron lifetime shorter than the age of the system, the distance that a particle can travel is determined by its synchrotron lifetime rather than the system age. In the vicinity of the pulsar where the magnetic field is strong, and the particles are transported mainly by energy-independent advection, the PWN size in different energy bands should be proportional to the synchrotron lifetime of the electrons (e.g., \citet{Tang12}). This is demonstrated by the changing nebula size in energy observed with \nustar{}. A synchrotron lifetime $t_{sync}$ can be defined as a time scale that an electron with Lorentz factor $\gamma$ loses all of its energy $E_{\gamma}$ via synchrotron radiation in magnetic field strength B. A synchrotron spectrum of a single electron is highly peaked around its critical frequency $\nu_{crit}(\gamma) \propto BE_{\gamma}^2$. A rough estimation of a synchrotron lifetime using this information yields
\begin{equation}
    t_{sync}=\frac{E_{\gamma}}{P_{synch}} \sim \frac{\gamma mc^2}{\gamma^2B^2}\sim \frac{1}{ \sqrt{B^3\nu_{crit}(\gamma)}}.
\label{eq:tsync}
\end{equation}
Assuming a constant average magnetic field and advection velocity in the region, the ratio of a synchrotron lifetime between particles emitting 3 keV photons and those emitting 6 keV photons is calculated as $\sqrt{6\textrm{ keV}/3\textrm{ keV}}=1.4$. This ratio is indeed comparable to the ratio of the nebula size in two different energy bands: FWHM(3$-$6 keV)/FWHM(6$-$20 keV)$=26.5\arcsec/15.2\arcsec=1.3$. 

Comparing the nebula sizes in the two energy bands also allows placing an upper limit of the magnetic field inside the compact nebula. The inner nebula detected by \nustar{} is located well outside the termination shock (2$-$3 smaller than the torii $\sim 10\arcsec$ (\citet{Etten08})), where the advection velocity can be approximated to the overall PWN expansion velocity (e.g., \citet{Porth16}). The expansion velocity of the Dragonfly has not been measured, yet some other PWNe were estimated to expand at $\sim$1,000 km/s (e.g., \citet{Porth16}, \citet{Reynolds18}, and \citet{Vorster13}). Eq. (6) in \citet{Reynolds18} gives the time it takes for an electron to lose half its energy via synchrotron radiation ($t_{1/2}$). For example, an electron that was emitting 12 keV photons $t_{1/2}$ years ago has cooled down by now to emit 3 keV photons ($\nu_{crit} \propto E^2$). Assuming that this electron traveled from the edge of the hard-band nebula (FWHM = $15.2\arcsec$) to the edge of the soft-band nebula (FWHM = $26.5\arcsec$) at velocity $v_{ad}=$1,000 km/s, the constant average magnetic field inside the compact nebula yields $B=24d_{3.5}^{-3/2}\mu$G. Compared with the 2.7 $\mu$G in the outer nebula, the much stronger magnetic field for the outer nebula was anticipated from the compact size of the hard X-ray inner nebula. This magnetic field estimate is consistent with the inner nebula magnetic field estimated by \citet{Etten08} ($\sim$ 20 $\mu$G assuming a dipolar field) and \citet{Jin23} ($\sim$ 22 $\mu$G assuming equipartition between the magnetic field and particle energy).

\section{Summary} \label{sec:summary}

As part of our \nustar{} observational campaign of energetic PWNe, we studied the X-ray properties of the Dragonfly PWN and its viability as a leptonic PeVatron. Our \nustar{} observation detected a compact (r $= 1\arcmin$) inner nebula of the Dragonfly in $3-20$ keV. The size of this nebula decreases at higher energies, indicating synchrotron burnoff in a strong ($\sim 24$ $\mu$G) magnetic field near its pulsar \psr{}. The large diffuse outer nebula of the Dragonfly is observed in soft X-ray ($\sim 10\arcmin$) and radio ($\sim 2\arcmin$). We used these outer nebula spectra along with the TeV spectrum of \ehwc{} to model the MW SED of the Dragonfly. The dynamical model yields the maximum particle energy of 1.4 PeV, and a low magnetic field (2.7 $\mu$G) averaged over the outer nebula in contrast to the high magnetic field in the inner nebula. At a nominal distance of 3.5 kpc, this 16-kyr-old PWN was found to have collided with the SNR reverse shock 4 kyrs ago. The highly asymmetric and energy-dependent morphology of the Dragonfly implies a fast proper motion of its pulsar ($\sim$ 1,000 km/s) and/or inhomogeneity in the ISM that initiated an asymmetric SNR $-$ PWN interaction. Our future work will investigate these scenarios and provide a deeper understanding of particle transport in such an evolved system using radio, X-ray, and gamma-ray observations.

The Dragonfly shares common features with other PWNe in our \nustar{} observational campaigns with the maximum particle energies above 1 PeV $-$ the Eel (\citet{Burgess22}) and Boomerang (Pope et al. submitted to ApJ). These features include a compact hard X-ray inner nebula undergoing synchrotron burnoff, a large diffuse outer nebula in lower energy, and an absence of a GeV nebula. Opposite patterns are observed in two of our target PWNe, the K3 (\citet{K3}) and Rabbit (\citet{Rabbit}). These PWNe exhibit extended hard X-ray nebulae without a sign of synchrotron burnoff, energy-insensitive morphologies, and bright GeV nebulae. The best-fit multi-zone models of the two PWNe yield the maximum particle energies below 1 PeV, while the PWNe are invisible to the current UHE observatories. The next-generation UHE observatories in the southern hemisphere (SWGO and CTAO$-$South) will enable us to study the true energetics of the PWNe and its relation to the MW observations.

\begin{acknowledgments}
We thank Mattia Di Mauro for providing the GeV upper limits. We acknowledge Ruo-Yu Shang, Eric Gotthelf, and Jordan Eagle for their helpful discussions. We thank the referee for carefully reading our manuscript and providing valuable comments. Support for this work was partially provided by NASA through \nustar{} Cycle 6 Guest Observer Program grant NNH19ZDA001N. HA acknowledges support from the National Research Foundation of Korea (NRF) grant funded by the Korean Government (MSIT) (NRF-2023R1A2C1002718).
\end{acknowledgments}

\vspace{5mm}
\facilities{\nustar{}, \chandra{}, XMM-Newton, \fermi{}-LAT}
\software{NuSTARDAS (v2.0.0), CIAO (4.13; \citet{ciao}), SAS (v20.0.0; \citet{SAS}), Fermipy (v1.0.1; \citet{fermipy}), HEASoft (\citet{HEAsoft}), HENDRICS (7.0; \citet{Bachetti18}), Stingray (v1.1; \citet{Huppenkothen19}), Xspec (\citet{xspec}), Naima (\citet{Naima})}

\bibliography{main}{}
\bibliographystyle{aasjournal}

\end{document}